\documentclass[preprint]{aastex}
\usepackage{natbib}

\def\kms{km s$^{-1}$}
   
\def\csn{$\chi^2_\nu$}
\def\emax{$E_{max}$}

\def\al{$\alpha$}

\def\snei{{\it nei }}

\def\vshock{{\it vpshock }}
\def\vpshock{{\it vpshock }}

\def\srcut{{\it srcut }} 
\def\sresc{{\it sresc }}

\def\rolloff{{\it rolloff}}
\def\norm{{\it norm}}

\def\power{ power-law }

\def\rosat{{\sl ROSAT}}

\def\asca{{\sl ASCA}}
\def\sis{ SIS }
\def\gis{ GIS }
\def\rxte{{\sl RXTE}}

\def\pca{ PCA }

\def\osse{{\sl OSSE}}

\def\cangaroo{{\sl CANGAROO}}

\shortauthors{Dyer et al.}
\shorttitle{Thermal and Nonthermal X-Rays:SN~1006}
                                                       
\begin{document}

\title{Separating Thermal and Nonthermal X-Rays in Supernova Remnants
I: Total Fits to SN~1006 AD}

\author{K.K. Dyer, S.P. Reynolds, K.J. Borkowski} \affil{North
Carolina State University, Physics Dept. Box 8202, Raleigh NC
27695-8202}

\author{G.E. Allen} \affil{MIT Center for Space Research NE80-6015, 70
Vassar St Cambridge, MA 02139}

\author{R. Petre} \affil{NASA's GSFC, LHEA Code 666, Greenbelt MD
20771} \email{Kristy\_Dyer@ncsu.edu}


\begin{abstract}

The remnant of SN~1006 has an X-ray spectrum dominated by nonthermal
emission, and pre-ASCA observations were well described by a
synchrotron calculation with electron energies limited by escape.  
We
describe the results of a much more stringent test: fitting spatially
integrated \asca \gis (0.6 -- 8 keV) and \rxte~\pca~(3 -- 10 keV) data
with a combination of the synchrotron model \sresc newly ported to XSPEC 
and a new thermal shock model \vshock.  The new model can describe the continuum emission above 2 keV well, 
in both spatial distribution and spectrum.  We find that the
emission is dominantly nonthermal, with a small but noticeable thermal component: 
Mg and Si are clearly
visible in the integrated spectrum.  The synchrotron component rolls
off smoothly from the extrapolated radio spectrum, with a
characteristic rolloff frequency of $3.1 \times 10^{17}$ Hz, at which
the spectrum has dropped about a factor of 6 below a powerlaw
extrapolation from the radio spectrum.  Comparison of TeV observations with new TeV model images and
spectra based on the X-ray model fits gives a mean post-shock magnetic
field strength of about 9 $\mu G$, implying (for a compression ratio of 4)
an upstream magnetic
field of 3 $\mu G$, and fixing the current energy content in relativistic
electrons at about $7 \times 10^{48}$ erg, resulting in a current
electron-acceleration efficiency of about 5\%. 
This total energy is about 100 times the energy in the magnetic field.  
The X-ray fit also
implies that electrons escape ahead of the shock above an energy of
about 30 TeV.  This escape could result from an absence of scattering
magnetohydrodynamic waves above a wavelength of about $10^{17}$ cm.
Our results indicate that joint thermal and nonthermal fitting, using
sophisticated models, will be required for analysis of most
supernova-remnant X-ray data in the future.
\bigskip

\end{abstract}

\bigskip

\section{Introduction}
\label{sec:introduction}


Since supernova shocks are one of the few mechanisms known to be capable
of providing adequate energy to supply the pool of Galactic cosmic rays,
supernova remnants (SNRs) have long been suspected as the primary
site of Galactic cosmic ray acceleration, at least up to the slight steepening in
the cosmic-ray spectrum at a few $10^{15}$ eV, known as the ``knee.''
Direct evidence for energetic particles comes from radio observations of
synchrotron emission from 1--10 GeV electrons. However, in the cosmic rays
observed at Earth, at a few GeV electrons are about 50 times less
numerous than cosmic-ray ions, whose spectrum is an unbroken power law
from $10^9$ eV to 10$^{15}$ eV. These considerations raise two questions:
1) Do SNRs accelerate ions? 2) Are they capable of accelerating {\it any}
particles to energies of 10$^{15}$ eV?  This paper addresses the second
question by demonstrating the presence of electrons of energies of hundreds
of TeV in the remnant of SN 1006 AD.

High-energy electrons can produce X-rays via nonthermal bremsstrahlung
and synchrotron radiation, and gamma-rays from inverse-Compton
upscattering of any photons present.  Relativistic protons can produce
$\gamma$-rays from the decay of $\pi^0$ particles produced in inelastic
collisions with background gas.  These processes have recently been
studied in detail by \citet{Sturner97}, \citet*{Gaisser98} and
\citet{Baring99}.  The analysis of X-ray and gamma-ray observations of
shell supernova remnants may give direct evidence bearing on both
questions above.  In this paper we will interpret observed X-rays and TeV
gamma-rays as primarily synchrotron radiation and inverse-Compton
upscattered cosmic microwave background radiation, respectively, and will
obtain maximum electron energies and electron shock acceleration
efficiencies -- crucial information for understanding shock
acceleration in general, and the origin of Galactic cosmic rays.

The earliest evidence of nonthermal X-ray emission in a shell supernova
remnant came from the featureless spectrum of SN~1006 AD 
(G327.6--1.4) \citep{Becker80}, 
explained as the loss-steepened extrapolation
of the radio synchrotron spectrum by \citet{Reynolds81}. However, early
data were poor and the models were simplistic.  Thermal models by
\citet*{Hamilton86} seemed able to produce a featureless X-ray spectrum,
given a high degree of elemental stratification in ejecta. However,
observations by \asca~ \citep{Koyama95} showed unmistakable evidence for
nonthermal emission in the rims, along with thermal, line-dominated
emission in the interior.  \citet{Reynolds96} demonstrated that detailed,
self-consistent synchrotron models could be constructed which adequately
described the pre-\asca~ integrated spectrum of SN~1006 as the diminishing
extension of the radio spectrum, reproducing the slope reported
by \citet{Koyama95} in the rim.  While \citet{Laming98} proposed a
modified thermal bremsstrahlung model based on the model of
\citet{Hamilton86}, even the author concluded that it is not an
appropriate fit for SN~1006.

Since then, several more shell SNRs have been shown to have nonthermal,
non-plerionic emission.  In addition to  SN~1006 \citep{Koyama95},
G347.3-0.5 (RJ J1713.7-3946) \citep{Koyama97, Muraishi00,Slane99} has
nearly featureless X-ray spectra clearly dominated by nonthermal
emission.  Some remnants, while not dominated by nonthermal emission in
the \asca~band, show other evidence pointing to its presence. \osse~
observations of Cassiopeia A at 400-1250 keV \citep{The96} are well
described by a broken power law, steepening to higher energies
\citep{Allen97a}. 
RCW 86 \citep*{Vink97} shows
anomalously low abundances when fit with thermal models.  When the fit
includes a synchrotron model the abundances fall within expected ranges
\citep{Borkowski99}.
 
An important consequence of synchrotron X-ray emission was pointed out by
\citet{Pohl96} and \citet{MastdeJ96}.  Electrons capable of producing keV
synchrotron photons in magnetic fields of a few microgauss will also
produce high-energy photons by inverse-Compton scattering any photon
fields.  In particular, inverse-Compton upscattering of
cosmic-microwave-background photons can produce TeV gamma rays, and
\citet{Pohl96} explicitly predicted a range for the expected TeV flux of 
SN~1006 (subsequently found by the CANGAROO
air \v Cerenkov telescope). The predicted flux depends only on the electron
distribution, so it can be used, in conjunction with observed synchrotron
fluxes, to deduce a mean magnetic-field strength in a remnant.  
Subsequent extensive calculations by \citet{Sturner97}, \citet{Gaisser98},
and \citet{Baring99} included this effect.
 
However, nonthermal X-ray emission from shell supernova remnants could be
due to bremsstrahlung from nonrelativistic electrons with energies of a
few tens of keV \citep{Asvarov90} rather than synchrotron radiation
from electrons with energies of hundreds of TeV.  Careful models (e.g.,
\citealp{Baring99} as well as \citealp{Reynolds96}) show that in neither
case are straight power-laws to be expected. Synchrotron spectra should
all be steepening (convex up) while bremsstrahlung spectra, resulting
from the lowest energy cosmic ray electrons departing from the Maxwellian
tail, ought to be flattening (concave up). \citet{Allen97a} showed that the
\asca~spectrum from SN~1006 steepens to higher energies, and the
discovery of the predicted TeV gamma rays by the \cangaroo\ collaboration
\citep{Tanimori98} provided strong evidence that the nonthermal X-ray
emission in SN~1006 is in fact synchrotron radiation.  However,
nonthermal bremsstrahlung is expected to be important for some remnants,
and as more observations become available above 10 keV, it will be
necessary to discriminate between synchrotron and nonthermal
bremsstrahlung.  (While most analyses, e.g., \citet{MastdeJ96}, conclude
that the TeV emission from SN~1006 is inverse-Compton, \citet{Aharonian99}
and \citet*{Berezhko99} assert that the $\pi^0$-decay mechanism may be
partly responsible.)

\begin{figure} 
\plotone{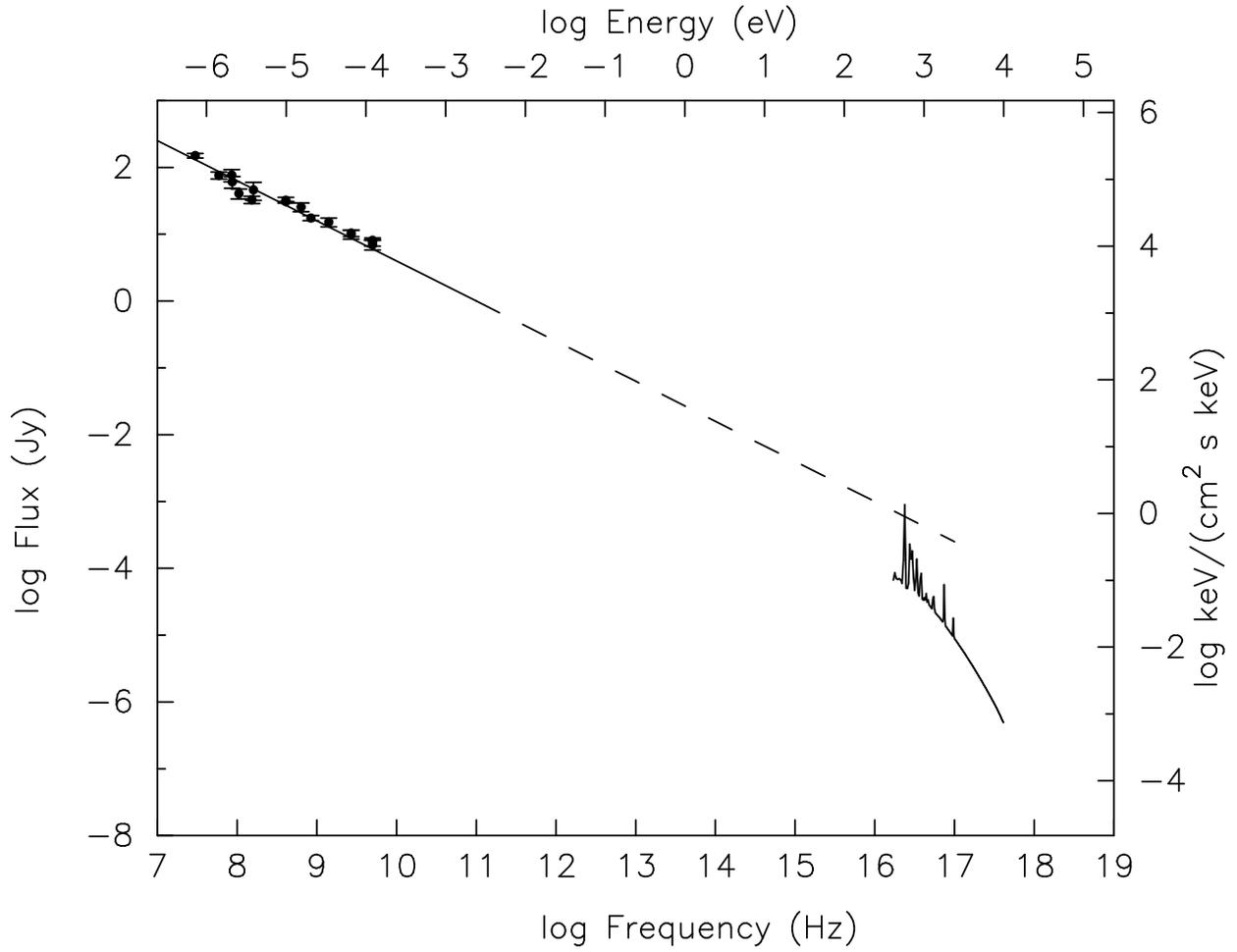}
\caption{The radio flux measurement with a power law spectrum extrapolated to X-ray
energies. The X-rays are the unfolded \asca~spectrum, shown only for comparison. \label{fluxes}} \end{figure}

Despite the strong evidence for curvature, a straight power law is still
commonly regarded as a defining characteristic of X-ray synchrotron
radiation in shell supernova remnants, as it is in Crab-like supernova
remnants and active galactic nuclei.  While power laws will certainly be a
better description of synchrotron emission produced by the particle
spectra expected in SNR shock acceleration than quasi-thermal exponential
continua, all SNR X-ray synchrotron continua are expected to be
substantially curved, even over a relatively narrow energy range.  
However, in almost all published studies, nonthermal SNR X-ray
emission has been described by a power law. This purely
phenomenological treatment, while characterizing the emission, fails
to explain its origin. Power laws cannot continue forever rising as
the frequency drops, a fact camouflaged by the rapidly dropping
response of X-ray instruments at low energies. Broken power laws can
avoid this ``infrared catastrophe,'' but they still duck the question
of the origin of the emission and its relation to the lower-energy
synchrotron spectrum. Since no Galactic supernova remnant has X-ray
emission bright enough to lie on the unbroken extrapolation of the
radio spectrum \citep{Reynolds99}, extrapolating from radio
measurements will grossly over-predict the X-ray flux, as demonstrated
in Figure \ref{fluxes}. The spectrum must drop significantly below the
observed power law in the radio to be consistent with the X-ray flux
measurements. A good physical model must both fit this turnover and
explain the mechanism causing the spectral curvature.

Some spectral modeling of SNR synchrotron X-rays has been done using
other descriptions.  While a power law represents a spectrum with no
curvature, the maximum curvature likely to result from shock
acceleration comes from an electron spectrum with an exponential
cutoff.  In the approximation in which each electron radiates all its
energy at the peak frequency of the single-particle synchrotron
emissivity (the ``$\delta$-function approximation''), this implies a
photon spectrum $S(E) \propto
\exp((E/E_m)^{-1/2})$. \citet{Aharonian99} modeled synchrotron
emission from SN~1006 using this expression, with a constant maximum
energy $E_m$.  As was demonstrated in \citet{Reynolds98} the
$\delta$-function approximation is not accurate well above the
turnover energy. In addition, the electron distribution will vary
throughout the remnant.  A convolution of the electron synchrotron
emissivity with the calculated electron distribution at each location
in the remnant is required. The model described in this paper includes
the full electron emissivity.

It is important to understand and accurately characterize the synchrotron
emission in SNRs. If the emission mechanism in those SNRs with clear
nonthermal continua is synchrotron radiation, then many SNRs could have
varying amounts of synchrotron X-ray emission, contributing part of the
continuum emission alongside the thermal lines and continuum. As
demonstrated by RCW 86 \citep{Borkowski00a}, such an unmodeled component stymies
thermal fits by preventing accurate measurements of shock temperatures and
elemental abundances. Understanding either the thermal or nonthermal
X-ray emission from SNRs requires appropriate modeling of both components.
This paper applies a model that assumes Sedov dynamics and a power-law
distribution of electrons up to a maximum energy \emax, which can vary
with time and with location in the remnant.  In general, the maximum
energy attained by electrons in the shock acceleration process could be
limited by one of several factors --- electrons above this energy could
escape from the remnant, the remnant could be so young it has not had time
to accelerate particles beyond some \emax, or \emax~could be the energy at
which radiative losses balance further acceleration.  Normally, a
remnant's spectrum would first be limited by age, and then by radiative
losses, unless escape prevented either of the other limits from being
reached.

In this paper we analyze new \rxte \ observations, along with archival
\asca \ observations, in Section \ref{sec:observations}. We discuss the
escape model, {\it sresc}, in Section \ref{sec:model}, along with our new
gamma-ray modeling, and compare \sresc~to thermal models and a
thermal+power law model in Section \ref{sec:results}. In Section
\ref{sec:discuss}, we discuss the 
elemental abundances we infer from the joint thermal-nonthermal model,
and compare them to predictions by \citet*{Nomoto84} and
\citet{Iwamoto99} and discuss the implications of our results for the efficiency
of electron acceleration.  We summarize our conclusions in Section
\ref{sec:conclusions}.

\section{Observations}
\label{sec:observations}

\subsection{RXTE}

On 18--19 February 1996, SN~1006 was observed for 21 kiloseconds with the
Proportional Counter Array (PCA) of the \rxte~ satellite.  A summary of
\rxte \ observations can be found in Table \ref{rxte}. The \pca~is a
spectrophotometer comprised of an array of five co-aligned proportional
counter units that are mechanically collimated to have a field-of-view of
$1\arcdeg$ at full width, half maximum.  The array is sensitive to 2--60 keV photons and has a
maximum collecting area of about 7000 cm$^2$ \citep{Jahoda96}.  The
instrument was pointed at a location on the bright north-eastern rim
($\alpha_{2000} = 15^{\rm h}~4\fm0$, $\delta_{2000} =
-41\arcdeg~48\arcmin$).  The angular diameter of SN~1006 is small enough
(about $30\arcmin$) that the entire remnant was in the field-of-view of
the instrument.  The total event-detection rate during this observation is
$26.43 \pm 0.06$ count s$^{-1}$ in the range 3--10 keV.  The background
spectrum of the PCA data was estimated using version 1.5 of the program
{\it pcabackest}.  This version includes estimates of the charged-particle
and diffuse cosmic X-ray backgrounds based on the ``VLE'' count rate
during observations of ``blank-sky'' regions and an estimate of the
background associated with the decay of radioactive material that is
activated when the spacecraft passes through the South Atlantic Anomaly.  
The count rate of the estimated background is $15.46 \pm 0.06$ count
s$^{-1}$ (3--10 keV).  Time intervals during which (1) one or more of the
five proportional counter units is off, (2) SN~1006 $< 10\arcdeg$ above
the limb of the Earth, (3) the background model is not well defined, and
(4) the nominal pointing direction of the detectors $> 0\fdg02$ from the
specified pointing direction are excluded from the present analysis.  The
count-rate attributed to SN~1006 during the 6.6 kiloseconds of the observation that
satisfies these four criteria is $11.0 \pm 0.1$ count s$^{-1}$.

During the proposal phase we were concerned with possible contamination
from Lupus and special effort was made to obtain off-source spectra in
addition to spectra from SN~1006 AD. Lupus, a thermal source, is not
expected to produce significant emission above 3 keV and preliminary
examination of \rxte~data from Lupus reveal spectra very similar to the
cosmic X-ray background. This data has also been analyzed by Allen \citep[][Allen et al. 2000, in preparation]{Allen97a}.

\subsection{ASCA}

SN~1006 has been observed several times with \asca~in the performance
verification phase and later cycles. A summary of the \asca~observations
used in this paper can be found in Table \ref{asca}. Backgrounds for \sis
were obtained from off-source portions of the same chip. Backgrounds for
\gis were taken from the 3 November 1999 dark earth event files.
Since SN~1006 is located in the galactic anticenter, 15\arcdeg \ off the
plane, galactic emission should not contribute significantly to the
background.

In Section \ref{sec:north} we fit observations of the north limb of SN
1006, using areas from a single \sis chip plus the equivalent portions of
the \gis. In Section \ref{sec:full} we fit three \asca~ observations of
the whole remnant (\gis2 \& 3, for a total of six data sets) obtained from
1993-1996. Since \rxte~has no spatial resolution and a 1\arcdeg~field of
view only the full fields of the \asca \gis can be simultaneously fit with
\rxte~observations.

\section{The Escape Model}
\label{sec:model}

Two models were proposed for the pre-ASCA integrated
spectrum of SN~1006 in \citet{Reynolds96}, with different mechanisms causing the 
rolloff in the electron spectrum:  an age-limited model, which
required an upstream magnetic field of only 0.6 $\mu$Gauss, and
an escape-limited model, in which the upstream magnetic field
could be much larger.  The TeV detection \citep{Tanimori98}
constrains the mean magnetic field in the remnant, as described
below, to be of order 10 $\mu$Gauss, ruling out the age-limited
model.  (The third category of model, radiative-loss-limited,
 could not fit the pre-ASCA data.)
We therefore consider only the escape-limited model
as a description for SN~1006.

\subsection{Assumptions in the model} 

The model \sresc~presented in this paper and discussed in
\citet{Reynolds98} describes SNR emission with electron energies limited
by escape, which accounts correctly for variation of shock-acceleration
efficiency with obliquity and for post-shock radiative and adiabatic
losses. The model makes the following assumptions: 1) the remnant dynamics
are well described by the Sedov self-similarity solution (we eliminate the
singularity at the center of the Sedov solution by truncating the
emissivity at a radius where the velocity of the shock=10,000 \kms, an
ejection velocity appropriate to Type Ia supernovae); 2) the remnant is
expanding into a uniform medium with a uniform magnetic field, $B_1$; 3)
the density increases by a fixed factor $r$ downstream of the shock -- this
means that the magnetic field strength increases by a factor between 1 and
$r$ as the tangential component is compressed, depending on the angle
between the shock normal and the upstream magnetic field; 4) the postshock
magnetic field evolves by flux-freezing, with no turbulent amplification
after the initial compression; 5) the relativistic electrons have short
postshock diffusion lengths so that their density evolves in the same
way as the thermal-gas density.
Electrons that escape are allowed to diffuse
upstream from the shock along magnetic field lines, producing a faint
``halo''.

Recent models of nonlinear shock acceleration 
\citep[e.g., Berezhko Yelshin \& Ksenofontov][Ellison, Berezhko, \& Baring 2000]{Berezhko99,Baring99}
predict that
high efficiencies of particle acceleration mimic a lossy shock, with
overall compression ratios considerably higher than 4, and
correspondingly lower thermal-gas temperatures.  Since both the escape
model and the thermal models we use below predict only post-shock
quantities, they are largely independent of the compression ratio,
which is used only to infer pre-shock values of density and
magnetic-field strength bsed on the post-shock inferences.  The
thermal inferences have strictly no dependence on the compression
ratio, while the details of the fainter nonthermal emission in the
escape model (where the magnetic field is parallel to the shock
normal) are slightly dependent on $r$.  However, for the values of $r
\sim 7$ given by the nonlinear model of \citet{Berezhko99}, the
predictions are indistinguishable from $r = 4$.  In any case, the
reader should bear in mind that all statements about pre-shock
quantities assume some value of the compression ratio.

The model describes the drop-off in synchrotron emission compared to a
straight power-law.  As adapted for XSPEC \citep[the X-Ray Spectral Fitting Package,][]{Arnaud} it can be applied to
remnants with different total radio flux densities and spectral
indices.  Predicted images depend on the time at which the remnant was
assumed to enter the Sedov stage, and at the aspect angle between the
upstream magnetic field and the line of sight, but these parameters do
not affect the integrated spectrum.  In fact, the escape model can be
applied to any remnant which has been interacting with uniform-density
material with a uniform magnetic field for most of its lifetime, as
long as the electron spectrum is in fact cut off by escape, and not by
finite remnant age or by radiative losses.  This can only be checked
by reference to the expressions for those maximum energies, as given
for instance in \citet{Reynolds98}; we require $E_{\rm max}$ (escape)
$<$ min$(E_{\rm max} ({\rm age}), E_{\rm max}({\rm loss}))$.  The
assumption of Sedov dynamics, rather than (say) a late stage of
self-similar driven wave into uniform material, makes little
difference to the total spectrum.  Since most of the X-ray emission is
produced closely behind the shock, the detailed form of the drop-off
of synchrotron emissivity is not critical.  (It would be important in
modeling radio images, but not for the integrated spectrum.)

The window for applicability of the escape model is probably
intermediate in a remnant's life, between early times when the age
limitation is most restrictive (and in which a core-collapse remnant
may be interacting with stellar-wind material with a circumferential
magnetic field), and late times when radiative losses may be most
influential.

There are several reasons why electrons above some \emax~could escape the
remnant.  While an intuitively obvious limit for an electron's gyroradius
is the SNR diameter, it has been shown that other mechanisms dominate
before this limit is reached \citep{Lagage83}.  In our diffusive
acceleration picture, acceleration occurs as electrons scatter resonantly
from magnetohydrodynamic waves in the upstream and downstream fluids.  
While the downstream medium is likely to be highly turbulent, supporting
waves of all wavelengths, this may not be true of the upstream medium,
where waves are probably produced by the accelerated particles themselves.  
Without reference to a detailed mechanism, in the escape model we assume
that magnetohydrodynamic scattering waves are much weaker above some
wavelength $\lambda_{max}$ which corresponds to an energy \emax~of
electrons at this gyroradius. Since electrons with gyroradius $r_g$
scatter resonantly with waves of wavelengths $\lambda = 2\pi r_g =
2\pi (E/eB)$,
electrons will escape upstream once their energy reaches an $E_{max}$
given by
 
\begin{equation} E_{max} = \lambda_{\rm max}eB_1/4 \end{equation}

\noindent 
(after averaging over pitch angles) where $B_1$ is the upstream
magnetic field strength. Unlike the age and loss-limited
cases \citep{Reynolds98}, here \emax~does not change with time or
depend on the angle between the shock normal and upstream
magnetic field. This energy then corresponds to a
photon frequency 
\begin{equation} 
\nu_{\rm max} = 1.05 \times 10^{15}
\lambda_{17}^2 B_{\mu G}^3 \left( r \over 4 \right)
\end{equation} 

\noindent
where $\lambda_{17} \equiv \lambda_{\rm max}/10^{17}$ cm, $B_{\mu G}$ is the 
upstream magnetic field measured in microgauss and $r$
is the compression ratio.  Only one power of $r$ is involved since
$E_{\rm max}$ depends on $B_1$, and $B_2$ only enters because $\nu_{\rm max}$
reflects the particles with $E = E_{\rm max}$ radiating in the stronger
post-shock magnetic field.

The shape of the cutoff could depend on the detailed distribution of
waves, but is likely to be no steeper than an exponential. The escape
model assumes an electron distribution given by $N(E)=K E^{-p} {\rm
exp}(-E/E_{max})$ at the shock, and evolves it appropriately in the
remnant interior, including radiative (synchrotron and inverse-Compton)
and adiabatic expansion losses. At each point a synchrotron volume
emissivity is found by convolving the single-electron emissivity with this
distribution. Model images (shown in Figure \ref{steve}b) are formed by
integrating along a raster scan of lines of sight and the spectrum is
found by integrating the flux over the image at each frequency. The models
are discussed in more detail in \citet{Reynolds98}.

\subsection{The model \sresc}

The escape-limited model is particularly well suited to algorithmic
fitting processes like those in XSPEC, since the departure from a
power-law is described by a single parameter; a universal function
describes the shape of the rolloff, and the one parameter simply locates
the rolloff in frequency.  Note from Equation 2 that fixing $\nu_{\rm rolloff}$
does not fix $\lambda_{\rm max}$, $B_1$, or $r$, but only the combination
$\lambda_{\rm max}^2 B_1^3 \ r$.  The model can be found in XSPEC 11
under the name {\it sresc}. The
models are also expected to be available with the next release of CIAO,
the Chandra software. 

The model has three parameters: 
\begin{enumerate} 
\item the radio flux measurement (\norm) at 1 GHz 
\item $\alpha$, the radio spectral index
(flux density $\propto \nu^{-\alpha}$) 
\item the characteristic rolloff frequency (Hz), called \rolloff~
(for {\it sresc}, $\nu_{\rm rolloff} = 5.3 \nu_{\rm
max}$, for historical reasons;  $\nu_{\rm rolloff}$ is the frequency at
which the spectrum has dropped by a factor of $\sim 6$ below the
extrapolated radio power law; \citealp{Reynolds98}) \end{enumerate} The spectral index and 1 GHz
flux for SNRs are fixed by radio observations. For Galactic SNRs, they can
be found at Green's website\footnote{Green D.A., 1998, `A Catalogue of Galactic Supernova Remnants (1998 September version)', Mullard Radio Astronomy Observatory, Cambridge, United Kingdom (available on the World-Wide-Web at "http://www.mrao.cam.ac.uk/surveys/snrs/").}.

Finally it should be emphasized that \sresc is only designed to model
Sedov-phase remnants whose maximum electron energy is limited by electron
escape. It also presumes a remnant in or close
to the Sedov dynamical phase, which has been encountering a uniform
upstream medium with a constant magnetic field. It is not
appropriate for remnants in a highly inhomogeneous environment. It should not be
applied to core-collapse remnants still interacting with stellar-wind
material (although it could be appropriate for some core-collapse remnants 
which have expanded beyond the pre-supernova medium).  In addition, if
the compression ratio $r$ is very much greater than $4$, the model
will overestimate fluxes from the ``poles'' where the emission is faintest,
though the integrated spectrum will not be strongly affected.

Since \srcut~\citep[synchrotron radiation cut-off model,][]{Reynolds99} describes the simplest possible synchrotron source
with a minimum of assumptions, it can be applied to remnants of unknown
provenance or dynamical stage more safely.  Until an appropriate range of
models, including different dynamical stages and different external media,
is available, \srcut represent a significant improvement over fitting a
power law especially if the purpose
of the fit is to account for the nonthermal emission so that a thermal
model may be accurately fit. Models appropriate for core-collapse remnants, and describing age- and loss-limited cases have been calculated
\citep{Reynolds97,Reynolds99}, and will also be made available
in XSPEC when practicable.

\subsection{Uncertainties in radio flux and spectral index}
\label{sec:uncert} Since the \sresc model has only one free parameter
(\rolloff) with both the spectral index (\al) and the 1 GHz flux (\norm)
specified by radio measurements, errors in these values will have an
effect on the synchrotron fits, and therefore on values in the co-fitted
thermal model.  The sensitivity to \al~ is substantial, since a variation
of 0.05, over a frequency range between $10^9$ Hz and $10^{18}$ Hz (4
keV), results in an offset of a factor of 3.

In addition, since many interferometer radio maps may miss smoothly
distributed radio flux, care must be taken if the flux in spatially
distinct regions is to be used as input to the model. In Section
\ref{sec:north} we used \sresc to fit part of the remnant.  To obtain
the appropriate value of \norm~from our 1.34 GHz interferometer map
\citep{Reynolds86}, which is missing a good deal of flux, we added
a uniform flux across the image to raise the total flux in the remnant to the prediction
from the radio single dish measurement of 15.9 Jy, using a spectral
index \al$ = 0.6$ to scale the 1 GHz total flux of 19 Jy
\citep{Green98}. We measured the flux in the subregion, then scaled
the flux to the 1 GHz value for input into the \sresc model using
$S_{\nu1}/S_{\nu2}=(\nu_1/\nu_2)^{-\alpha}$.

\subsection{TeV gamma-ray consequences of \sresc}

In the \sresc model, the spatial dependence of the electron
distribution is entirely fixed by the assumption of Sedov dynamics
and the assumption that the relativistic-electron density tracks the thermal-gas
density, with the specification of $E_{\rm max}$ and $B_1$.  This
dependence is calculated taking into account synchrotron and cosmic
microwave background inverse-Compton losses as well as adiabatic
expansion losses.  Adding a radio flux density and spectral index fixes
the normalization of the distribution, given the value of $B_1$.
Given the electron distribution everywhere, it is a simple matter
to calculate the morphology and spectrum of TeV gamma-ray emission.
To extract this prediction, we have added the inverse-Compton kernel
described in \citet{Baring99}, basically the full Klein-Nishina result,
to {\it sresc}. Here it is natural to work in photon energies instead of
frequencies. The Klein-Nishina cross-section is given by \begin{equation}
\sigma_{\rm K-N}(\varepsilon_s,\,\gamma_{\rm e} ;
{\displaystyle{\varepsilon}_{\gamma}} )\; =\; {{2\pi r_0^2} \over
{\varepsilon_s\gamma_{\rm e}^2}}\, \biggl[ 2q\,\log_{\rm e}q +1+q-2q^2
+{{\Gamma^2 q^2(1-q)} \over {2(1+\Gamma q)}}\biggr]\ , \end{equation} with
$\Gamma =4\varepsilon_s\gamma_e$ being the parameter that governs the
importance (when $\Gamma\gtrsim 1$) of photon recoil and Klein-Nishina
effects, and with \begin{equation} q\;
=\;{{\displaystyle{\varepsilon}_{\gamma}} \over {4\varepsilon_s\gamma_{\rm
e} (\gamma_{\rm e} - {\displaystyle{\varepsilon}_{\gamma}}
 )}}\ , \quad 0\leq q\leq 1\ , \end{equation} where $\varepsilon_s m_{\rm
e} c^2$ is the initial photon energy,
${\displaystyle{\varepsilon}_{\gamma}} m_{\rm e} c^2$ is the upscattered
(final) photon energy, and $\gamma_{\rm e} =(E_{\rm e} + m_{\rm e} c^2)/
m_{\rm e} c^2$.  The constant $r_0=e^2/(m_{\rm e} c^2)$ is the classical
electron radius. 
This result assumes isotropic soft photon fields, the
case for the cosmic microwave background radiation.

The inverse Compton emissivity for isotropic photon fields is then (e.g.,
Blumenthal \& Gould 1970)
\begin{equation}
{dn_{\gamma}({\displaystyle{\varepsilon}_{\gamma}}
 ) \over dt} \; =\; c \int N_{\rm e}(\gamma_{\rm e})\, d\gamma_{\rm e}
  \int d\varepsilon_s n_{\gamma}(\varepsilon_s)\; \sigma_{\rm
K-N}(\varepsilon_s,\,\gamma_{\rm e} ;
{\displaystyle{\varepsilon}_{\gamma}} ) \ , \end{equation} where
$n_{\gamma}(\varepsilon_s)$ is the distribution of seed photons, and
$N_{\rm e}(\gamma_{\rm e})$ is the calculated electron energy distribution
at each point in the remnant \citep{Reynolds98}. This expression is used
with the blackbody photon distribution (expressed in terms of photon
energies) \begin{equation} n_{\gamma}(\varepsilon_s)\; =\; n_{\rm
BB}(\varepsilon_s) \equiv {{\varepsilon_s^2} \over {\pi^2\lambda_c^3}}\,
{{1} \over {e^{\varepsilon_s/\Theta} - 1}}\ , \quad \Theta = {{kT} \over
{m_{\rm e} c^2}} \ , \end{equation} with $T = 2.73$ K so that $\Theta
=4.6\times 10^{-10}$.  Here $\lambda_c=\hbar /(m_{\rm e} c)$ is the
Compton wavelength.

As with synchrotron emission, the inverse-Compton emissivity is calculated at each
point in the remnant, and integrated along a raster scan of lines of sight
to produce a gamma-ray image.  Images at different photon energies are
integrated over to produce a total spectrum.


\section{Results} 
\label{sec:results}

Since preliminary fits to the X-ray data were unable to constrain the
absorbing column density, we turned to past observations. While \citet{Koyama95} found an absorption of 1.8~$\times$~10$^{21}$ cm$^{-2}$ most other observations point to lower absorption. The Schweizer-Middleditch star seen through SN~1006 has a color excess $E(B-V)$
of 0.12 \citep*{Blair96}. From this, the empirical relation between color
excess and column density \citep{Gorenstein75} yields $N_H =
7.7 \times 10^{20}$
cm$^{-2}$. This is not in gross disagreement with previous \rosat~observations by \citet{Willingale96}, who found a column density of
$(3.9-5.7)~\times$ 10$^{20}$ cm$^{-2}$. BeppoSAX observations by \citet{Vink00} found an absorption of (8.8~$\pm$~0.5$)~\times~$10$^{20}$. For fitting purposes we adopt a
column density of 5~$\times$~10$^{20}$~cm$^{-2}$ for all models.
Details on the datasets used for
fitting are summarized in Table \ref{data}.
To account for instrumental differences and differences in background
subtractions we allowed multiplicative offsets between all the data sets.
The ratio of \rxte \pca to \asca \gis data is given in Table \ref{fits},
row 3. It varied between 0.63 and 0.70. 

In the cases where a thermal and nonthermal model were fit, where the
nonthermal model dominated the flux, the thermal model had little
leverage to determine some parameters accurately. In particular,
determining the absolute abundances (relative to hydrogen) requires
fixing the level of thermal continuum, which is particularly
difficult in the presence of synchrotron continuum emission.
We believe the obvious line features can be analyzed with somewhat
more confidence.  Formal fits to the data resulted in
overall high abundances, with correspondingly large
errors, but we believe the ratios between abundances (in particular
to silicon, whose K$\alpha$ emission is relatively well determined)
are more reliable than the absolute values obtained, and give
the ratios in Table \ref{fits}.    
For the \sresc+\vshock~fit in Section \ref{sec:sync+shock} we reversed the process. Rather than fixing hydrogen and helium, we fixed silicon, the most obvious line present in the data, allowing the emission measure and other lines to adjust. This produced abundance ratios very close to fits with hydrogen and helium fixed but allowed us to obtain more reliable error ranges, including upper limits on hydrogen and helium.  

SN~1006 presents a complicated mix of thermal and nonthermal spectra.
It has been known from some time that emission from the rim of SN~1006 is
almost entirely nonthermal \citep{Koyama95}. Since the results of
jointly fitting two models are rarely unique 
we constrained fits to the full remnant by first
fitting the north rim with a nonthermal model, either \power or \sresc
\footnote{Since \sresc was designed to describe the spectrum from the
remnant as a whole this is an approximation, though a reasonable one
since the limbs dominate the total spectrum. In a future paper we will use
spatially resolved versions of \sresc to describe subsections of the
SNR.}. We then used the parameters from the limb data to constrain the
nonthermal model used on the entire remnant.

\subsection{Model Fits to the North Limb} \label{sec:north}

\begin{figure} 
\plottwo{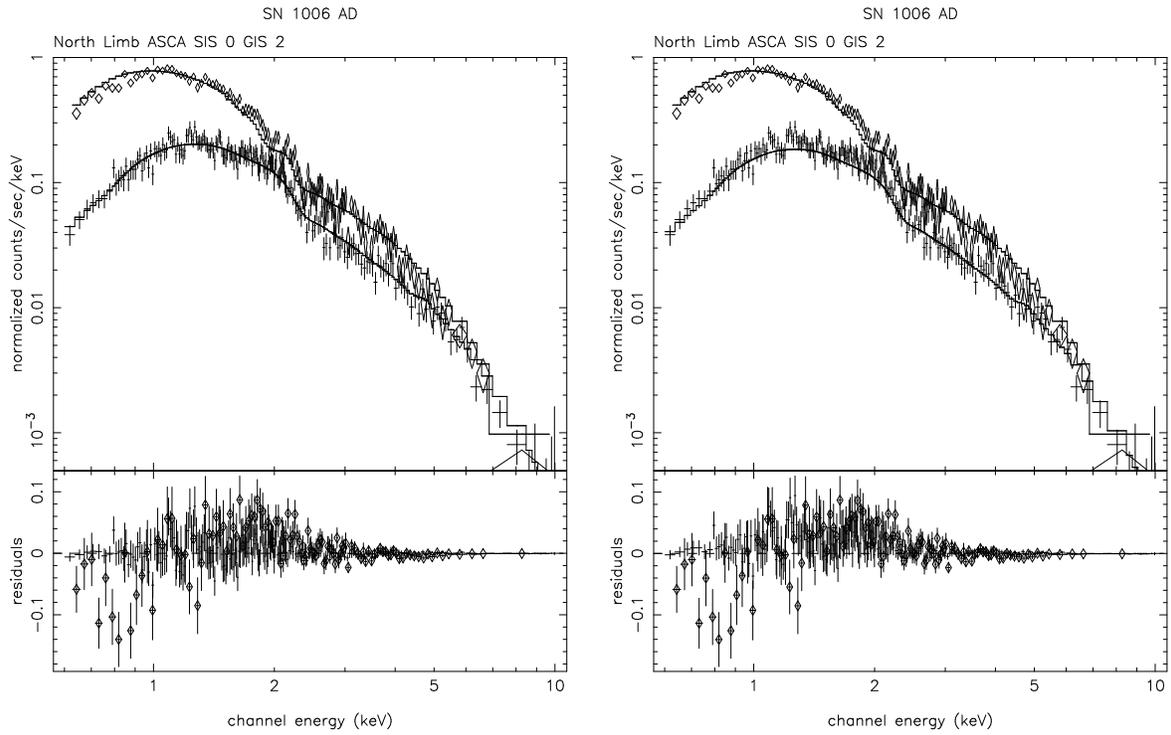}{f3b.ps} 
\caption{\asca \sis
0 \& 1 \gis 2 \& 3 (only \sis 0 (crosses) and \gis 2 (diamonds) are shown)
taken from the north rim, fit by a) \power and b) \sresc. Parameters for
the fits are given in \S \ref{sec:north}.\label{limb}} \end{figure}

We selected \asca \sis \ spectra obtained in the northeast observation 
(13 August
1993) from the single chips (\sis 0 chip 1, \sis \ 1 chip 3), with
corresponding regions from \gis \ 2 and \gis 3. We fit these featureless
spectra with a power law, as shown in Figure \ref{limb}a. We obtained a
photon index of 2.46 and a \norm~(photons keV$^{-1}$ cm$^{-2}$ s$^{-1}$ at 1 keV)
of $8.09 \times10^{-3}$,
 yielding a \csn = 0.91. All four datasets were allowed to vary by a
constant -- the ratios were \sis0:1.00 \sis1:0.91 \gis2:1.01 \gis3:0.76.
Using the energy range, column density, and line centroids from
\citet{Koyama95} we were able to replicate their results to within
1$\sigma$. We obtained $\Gamma$=2.50$^{2.34}_{2.63}$ (1 $\sigma$), in agreement with 
$\alpha=$1.95$\pm$0.2. (For power-law emission, using the explicit
positive convention for \al, the energy spectral index, $S_{\nu}\propto \nu^{-\alpha}$, and these
quantities are related by \al~+ 1 = $\Gamma$, where $\Gamma$ is the photon index.)

We then fit the same four data sets with \sresc, as shown in Figure
\ref{limb}b. We froze the spectral index \al~at 0.60 and froze \norm~
to the proportionate flux (as discussed in \S \ref{sec:uncert}) of 2.54
Jy. This left one free parameter, \rolloff=3.06$_{2.8}^{3.1}$$\times 10^{17}$~ Hz
yielding \csn=1.04. The ratios between the datasets were
\sis0:1.00 \sis1:1.02 \gis2:1.14 \gis3:0.88.

\subsection{Model Fits to Full Remnant} \label{sec:full}

Table \ref{fits} contains the results of the fits to the whole
remnant, for each of several thermal models as described below.
Included are the $\chi^2$ and the number of degrees of
freedom (DOF, row 1), the reduced $\chi^2$ (row 2), the multiplicative
offset used between the \rxte \pca and the \asca \gis (row 3), the
temperature found by each thermal model in units of keV
(row 4) and the ionization timescale $\tau \equiv n_e t$ in units of
s~cm$^{-3}$ (row 5). The fitted abundances are listed in rows 6-11,
given as $\case{<X/H>}{<X/H>_\sun}$ \citep[by number, not mass,
from][]{Grevesse89}, relative to silicon. Elements not listed could not be constrained by
the data and were left at solar. 
In row 13 the amount (\norm) of thermal/nonthermal component is given. For the powerlaw the
normalization is photons keV$^{-1}$ cm$^{-2}$ s$^{-1}$at 1 keV; for \sresc the
normalization is the 1 GHz radio flux. 
In row 14 the 0.6 -- 10.0 keV flux of the thermal and nonthermal model is given. 
Finally, in row 15, the
constraining nonthermal parameter is listed (power-law index for the
power law, and $\nu_{\rm rolloff}$ for \sresc~). This parameter is not
allowed to vary, but is taken from the corresponding fit in Section
\ref{sec:north}. Each model was fit to the full dataset (six \gis datasets + \rxte) but is shown in 
the figures only with a single \gis and the \rxte~ dataset for clarity. 
All errors given are 1 $\sigma$ errors. 

\subsubsection{Nonequlibrium Ionization Model}

The \asca-\rxte \ observations of SN~1006 were first fit with {\it NEI}, a
constant temperature non-equilibrium ionization model that describes an
impulsively heated uniform and homogeneous gas \citep*{Borkowski00b}, 
released with XSPEC v11. The \snei model, with elemental
abundances held fixed at solar, produced a \csn=3.27 (see Figure \ref{snei}). While
the \snei model represents the simplest non-equilibrium ionization
description of a thermal shock, we
do not expect it to be a good description since the physical case it
describes is only appropriate in limited cases where all emitting material
was shocked at the same time -- such as individual clumps of gas rapidly overrun
by the shock.

\subsubsection{Plane Shock Model}

\begin{figure} 
\plotone{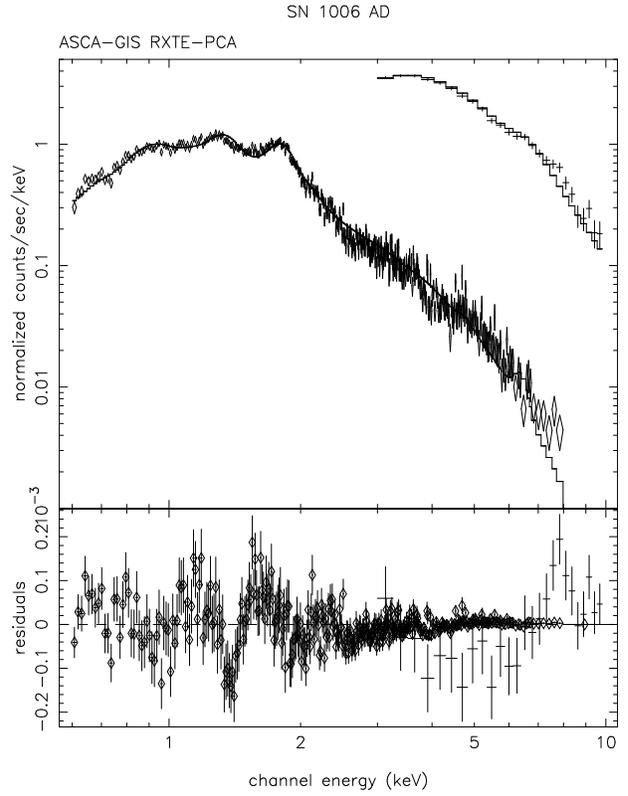} 
\caption{The \vshock model
applied to the full remnant \asca \gis (diamonds) and \rxte \pca
(crosses). Parameters listed in Table \ref{data}, column 3.\label{vshock}}
\end{figure}

The \vshock model is a plane-parallel shock model with varying abundances
\citep[][released with XSPEC 11]{Borkowski00b} including
nonequilibrium-ionization effects. It represents an improvement over the
\snei model by allowing a linear distribution of ionization timescale
$\tau \equiv n_e t$
vs.~emission measure between user-specified $\tau_{\rm lower}$ and
$\tau_{\rm upper}$.  Normally one will 
set $\tau_{\rm lower}$ equal to zero, for an ongoing shock, and we
have done so.  The abundances of O, Ne, Mg, Si,
S, and Fe were allowed to vary. We fit the \asca \gis + \rxte \pca data
with \vshock, shown in Figure \ref{vshock}. Were this the only information
we had on SN~1006 we would conclude that this is an acceptable description
of the data. However, the absence of spectral lines in the limb data (fit in Section 
\ref{sec:north}), indicates that a purely thermal explanation is not in fact adequate.

\subsubsection{Power Law + Plane Shock Model}

\begin{figure} 
\plotone{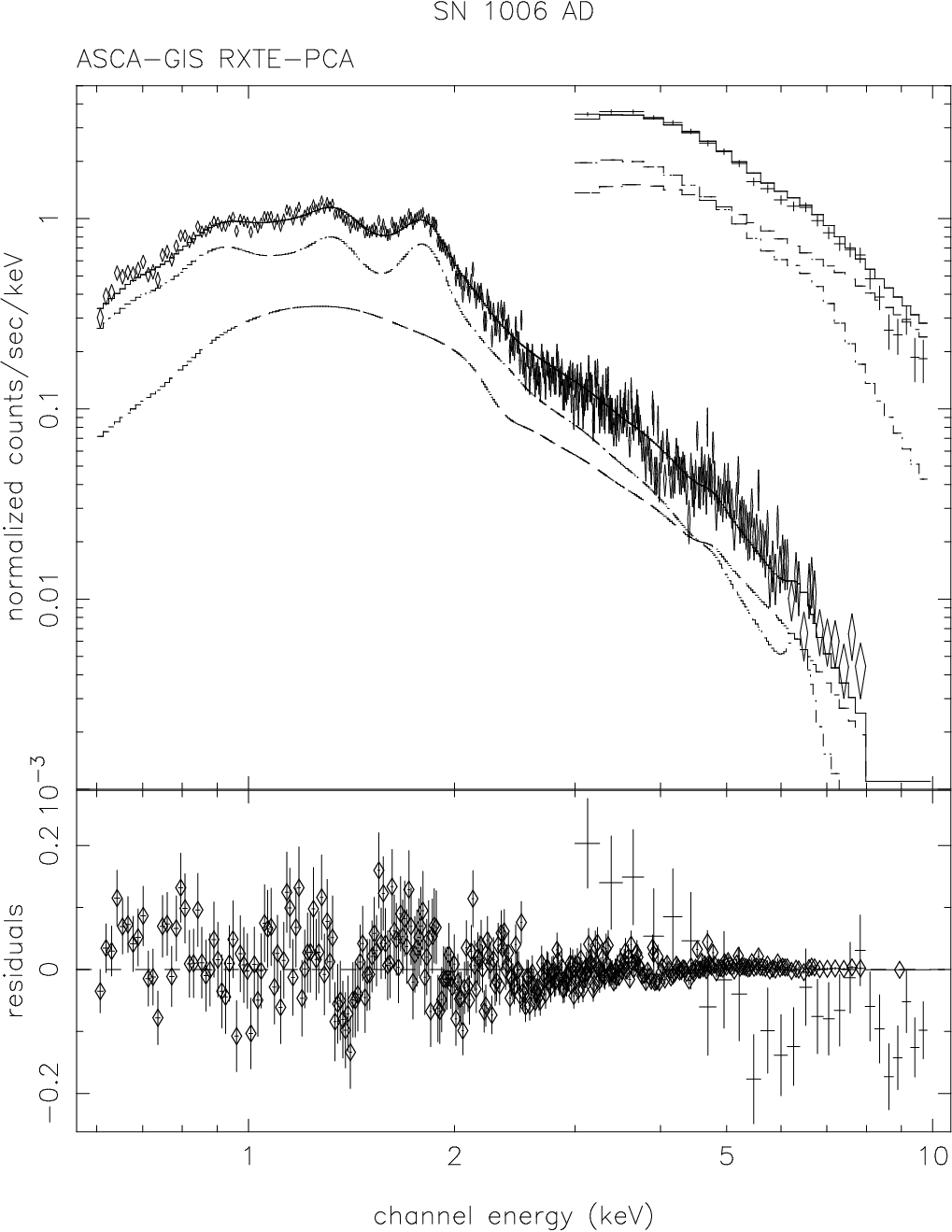} 
\caption{The \vshock + \power
model applied to the full remnant \asca \gis (diamonds) and \rxte \pca
(crosses). The solid line is the total model while the broken lines
represent the individual contributions: the smooth varying broken line is
\power while the \vshock shows promanent spectral lines. Parameters listed in Table
\ref{data}, column 4.\label{shockpower}} \end{figure}

It was impossible to find a unique separation between the \power and
\vshock components based only on the full-remnant spectrum. Lower fluxes of the \power
component were compensated for by the thermal model with lower temperatures and
higher abundances, with no outstanding fit.  Instead, we allowed the limb
data (\S \ref{sec:north}) to fix the power-law photon index. With that
photon index we set the highest possible \norm~that did not disagree with
the \rxte \pca data. We then froze the power law parameters and fit a
\vshock~model, allowing the abundances to vary. The fit is shown in Figure
\ref{shockpower}.

\subsubsection{Escape Limited Synchrotron + Plane Shock Model} 
\label{sec:sync+shock}

\begin{figure} 
\plotone{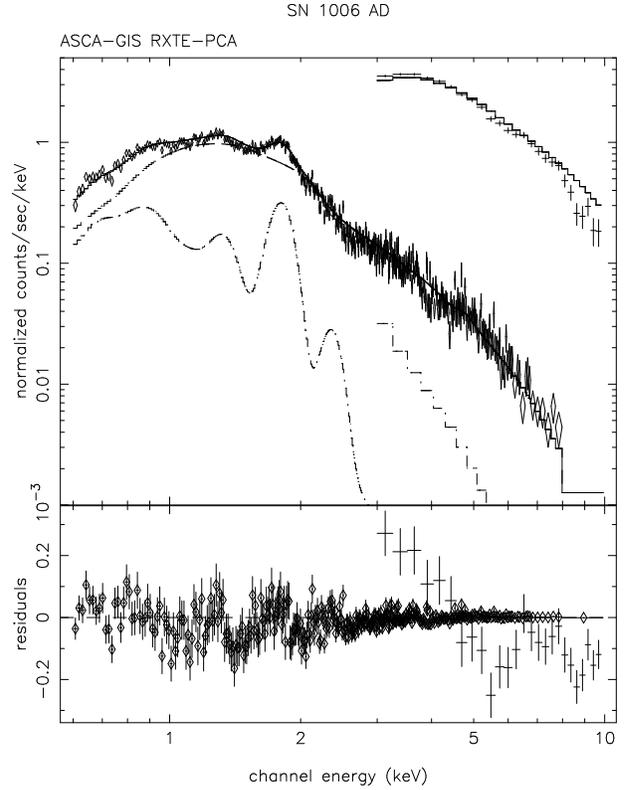} 
\caption{The \vshock + \sresc
model fit to whole remnant \asca \gis \ (diamonds)  and \rxte \pca
(crosses). The solid line is the total model while the broken lines
represent the individual contributions: the smooth varying broken line is
\sresc while the \vshock shows promanent spectral lines. 
Parameters listed in Table \ref{data}, column 5.\label{shockesc}} \end{figure}

We then fit an \sresc model plus \vshock. We took the parameters for
\sresc from the limb fits (\S \ref{sec:north}): \al= 0.60,
\rolloff=3.06$\times 10^{17}$ Hz, and \norm~(the 1 GHz flux)$=19$ 
Jy. This left no
free parameters in the \sresc portion of the fit. We then added a \vshock
component and allowed the elemental abundances to vary. The fit is shown
in Figure \ref{shockesc} and the results are listed in Table \ref{fits}. In the best fit \sresc+\vpshock model we found that 60\% of the flux was synchrotron emission and the thermal emission comprised the 40\% remaining X-ray flux. 
While the $\chi^2$ for \sresc + \vshock is identical to that of the \power + \vshock fit, there is only one \sresc + \vshock model, while as discussed above, many \power + \vshock models were possible. 

As discussed in Section \ref{sec:results}, we fixed silicon, the line with the most information in the spectrum, and fit all other parameters. The results are shown in the Table \ref{fits}, excepting hydrogen $0.00^{2.0E-3}$ and helium $0.00^{3.0E-3}$, for which it was only possible to find upper limits.

\begin{figure} 
\plottwo{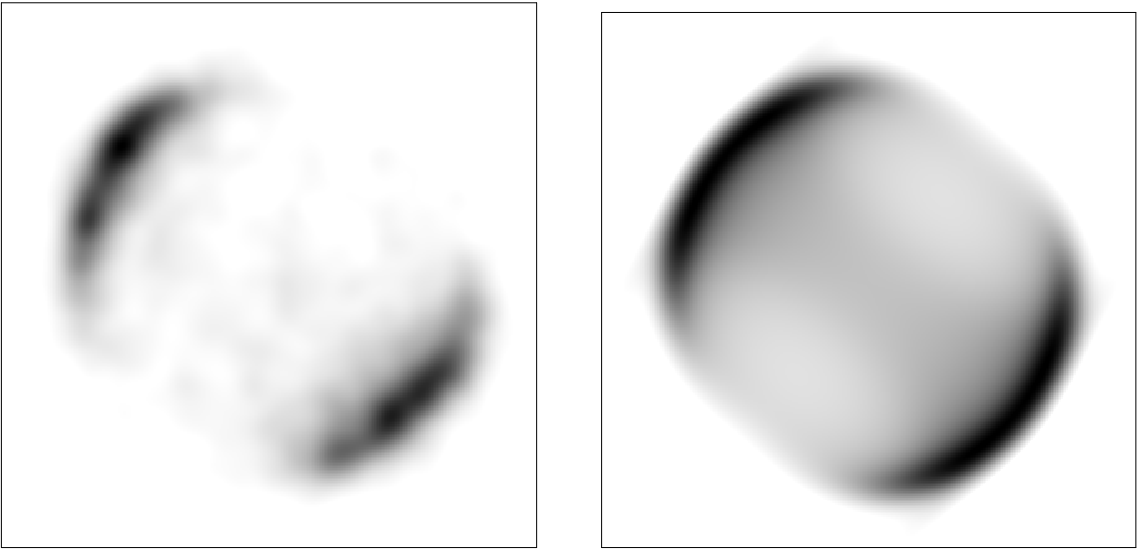}{f8b.ps} 
\caption{a) \gis image from
2-10 keV, energies dominated by synchrotron. b) The synchroton X-ray image
predicted by the \sresc model fitted parameters \rolloff = 3.0$\times$10$^{17}$, 
\al = 0.6, \norm
(1 GHz flux) = 19 Jy. We have assumed a uniform upstream magnetic field of
3 $\mu$G and a compression ratio of 4. 
Both images are convolved to 2\farcm \label{steve}} \end{figure}

The parameters of \sresc also yields spatial predictions. In Figure \ref{steve}a we show the 1993 August 19 \gis 2 data selected
from 2-10 keV, the portion of the spectrum where, according to
\sresc~models, the SNR is completely dominated by nonthermal emission. In Figure
\ref{steve}b we show the predicted image of the model at 2 keV generated
from the radio spectral index, the 1 GHz radio flux and the \sresc parameter, \rolloff. Both images have been convolved to
2\farcm~These images are strikingly similar, supporting the use of the synchrotron model.

\subsection{Gamma-ray Observations of North Limb}

Making use of the results of the above fits, we can also fit the 
TeV gamma-ray
observations of SN~1006 by \citet{Tanimori98}. The fitted values
for \sresc with \vshock produce a one-parameter family of predicted TeV
gamma-ray spectra, since two parameters (basically magnetic-field energy
density and relativistic-electron energy density, or $K$ and $B_1$) 
are required to set the
synchrotron normalization, and \norm~only fixes the product $K B^{1 + \alpha}$. 
The value
of $\nu_{\rm rolloff}$ from the \sresc fits fixes the product
$\lambda_{\rm max}^2 B_1^3$. Different choices of $B_1$, then, imply
different inferred values of $\lambda_{\rm max}$, otherwise unobservable,
and different values of the normalization constant $K$ in the electron
distribution.  However, each value of $K$ gives a different normalization
to the inverse-Compton spectrum which has no dependence on magnetic field.

\begin{figure} 
\plotone{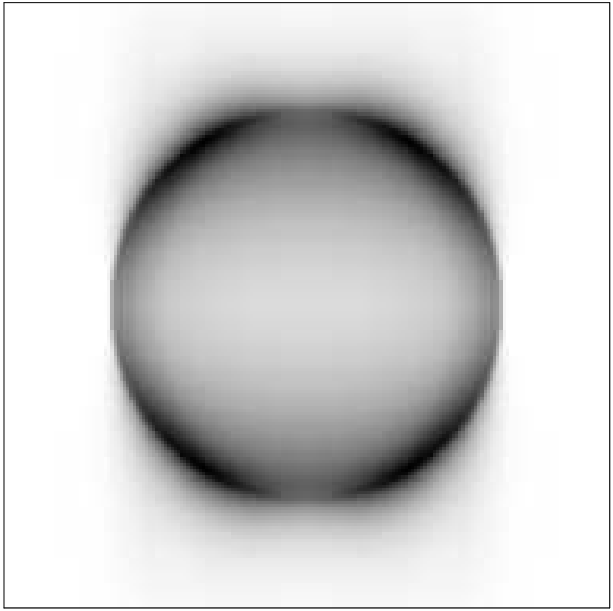}
\caption{The predicted image of gamma-ray emission at an
energy of 1 TeV. \label{gammaimage}} \end{figure}

The integrated fluxes of $(4.6\pm2.0)\times 10^{-12}$ photons cm$^{-2}$
s$^{-1}$ ($E\ge1.7\pm0.5$ TeV) and $(2.4\pm1.2)\times10^{-12}$
photons cm$^{-2}$ s$^{-1}$ ($E\ge3.0\pm0.9$ TeV) imply an energy
spectral index of 2.15. Using this value, the fluxes imply monochromatic
flux densities of $3.5\times10^{-38}$~erg~cm$^{-2}$~s$^{-1}$~Hz$^{-1}$
at $\nu=4.1\times10^{26}$~Hz, and $1.0\times10^{-38}$~erg~cm$^{-2}$~s$^{-1}$~Hz$^{-1}$ at 
$\nu=7.2\times~10^{26}$~Hz.  These fluxes are
plotted in Figure \ref{gammaspectra}, which also 
shows two gamma-ray spectra, for $B_1 = 3$ and
$5 \ \mu$gauss (and $r = 4$), 
constrained to have the same value of \norm, i.e., varying
in $K$.  The observed TeV gamma-ray flux then fixes the magnetic field,
along with $K$ and $\lambda_{\rm max}$.  A value of $B_1 = 3 \ \mu$gauss
describes the data well, and implies $\lambda_{\rm max} = 1.4 \times
10^{17}$ cm and $K = 5.9 \times 10^{-12}$ erg$^{p-1}$ cm$^{-3}$.  Figure
\ref{gammaimage} shows the predicted image of gamma-ray emission at an
energy of 1 TeV.  If $r > 4$, all values of $B_1$ are multiplied by $4/r$.

\section{Discussion} 
\label{sec:discuss}

\subsection{Thermal and nonthermal continuum}

As expected, the simple non-equilibrium ionization model, {\it nei}, with
solar abundances is a poor description to SN~1006. \vshock is a better fit
but knowing that parts of the remnant are dominated by nonthermal
emission we know that temperature and abundances change with the
addition of a nonthermal model.

The \power model was very difficult to constrain in conjunction with
thermal models. {\it vpshock}, and we suspect other thermal models as well, can
seemingly adapt to nearly any amount of \power: larger amounts of \power
cause lower temperatures, as \vshock contributes less of the continuum,
and higher abundances, as \vshock tries to account for line emission from
a lower thermal continuum.

The \sresc + \vshock described the \asca~ and \rxte~ observations very
well, at least as well as the best \vshock + \power. In addition, if
$\nu_{\rm \rolloff}$ is constrained from fits to the nonthermal dominated
limb observation, then remarkably \sresc has no free parameters -- it
uniquely specifies the synchrotron emission in the SNR.  This result
confirms that nonthermal X-ray emission can be well described by simple synchrotron model, and the synchrotron escape model, which has previously only been tested on integrated fluxes, provides a good description of the data.

High abundances of heavy elements in the \sresc + \vpshock model imply that
thermal emission is strongly dominated by supernova ejecta, which became obvious only after a proper separation of nonthermal and thermal components. Elemental abundances in SN~1006 ejecta are clearly of great interest, but 
spatially-integrated \gis~and \rxte~spectra used in this work allow only for 
the most basic interpretation of thermal spectra. 


The {\it vpshock} model is the best first approximation for modeling
X-ray spectra from shocks dominated by heavy element ejecta. Because this
model provides statistically and physically reasonable fits to spatially
integrated \gis~spectra, we briefly interpret our results. However, these are just preliminary results, a more detailed analysis of spatially-resolved
spectra should provide more secure answers. In this future analysis, one needs
to consider both the reverse and forward shock, instead of just one plane
shock considered here. This analysis should also be based on full hydrodynamical models with the realistic Type Ia ejecta structure \citep{Dwarkadas98}.



By freezing silicon, we found the best fit model required no hydrogen or helium with a 1-$\sigma$ upper limit of $\slantfrac{<H>}{<Si>}$=2$\times$10$^{-3}$ and $\slantfrac{<He>}{<Si>}$=3$\times$10$^{-3}$. We can use this to try and set an upper limit on the preshock density. 
From an upper limit on hydrogen of $\slantfrac{<H>}{<Si>}$=2$\times$10$^{-3}$ we obtain an upper limit on the preshock density of $0.014 (4/r)$ cm$^{-3}$. This is somewhat
less than $n_H$ of 0.05--0.1 cm$^{-1}$ (assuming $r=4$) estimated by \citet{Dwarkadas98} in their detailed analysis of SN~1006. However, our value is poorly constrained, since
it depends on determining the level of thermal continuum, which we cannot
do with certainty in the presence of much stronger nonthermal continuum.

But most emission in the \vpshock~model comes from heavy elements, presumably ejecta 
shocked by the reverse shock, and the temperature $T$ and ionization 
timescale $\tau$ in the last column of Table \ref{fits} most likely refer to the reverse 
shock. 
We can roughly estimate the time-averaged electron density 
$\langle n_e \rangle$ in the shocked 
heavy-element plasma by assuming that most X-ray emission comes from material 
shocked in the last 500 yr (approximately half of the remnant's age), which
gives $\langle n_e \rangle \sim \tau/500~{\rm yr} = 0.24$ cm$^{-3}$. This
order of magnitude estimate, together with abundances and normalization of
the \vpshock~model from Table \ref{fits}, allows us to arrive at the total mass of X-ray 
emitting ejecta of 0.71 M$_{\odot}$, a lot of it (0.45 M$_{\sun}$) in 
Fe. While the absolute numbers are uncertain, a high Fe mass is consistent
with the Type Ia SN progenitor.

\subsection{Elemental Abundances}

Extensive theoretical work has been carried out by \citet{Iwamoto99} 
and \citet{Nomoto84} predicting abundances expected from different supernova
types.  However it has proved quite difficult in practice to compare
theory directly to measured abundances in a SNR.

\begin{figure} 
\plottwo{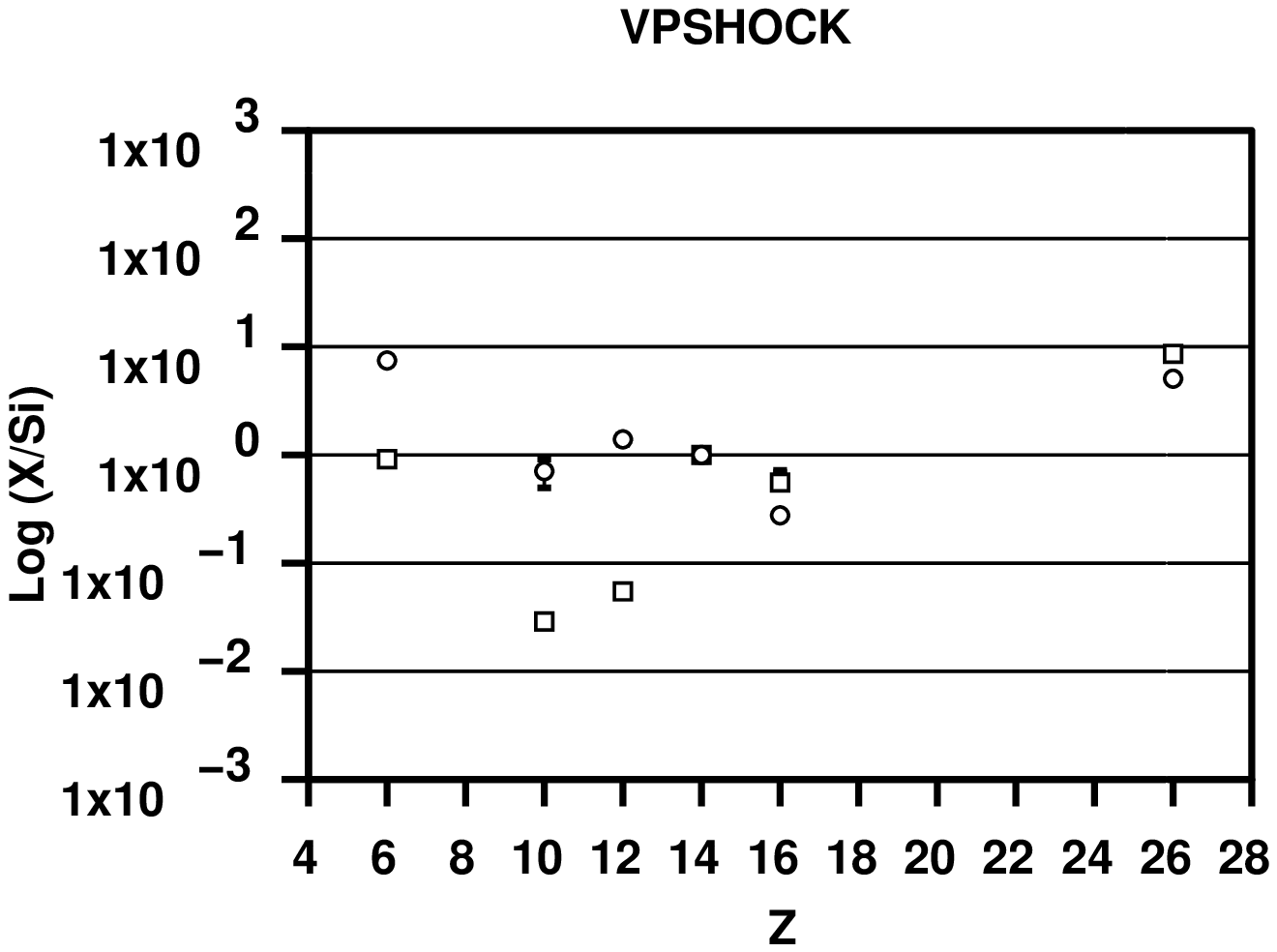}{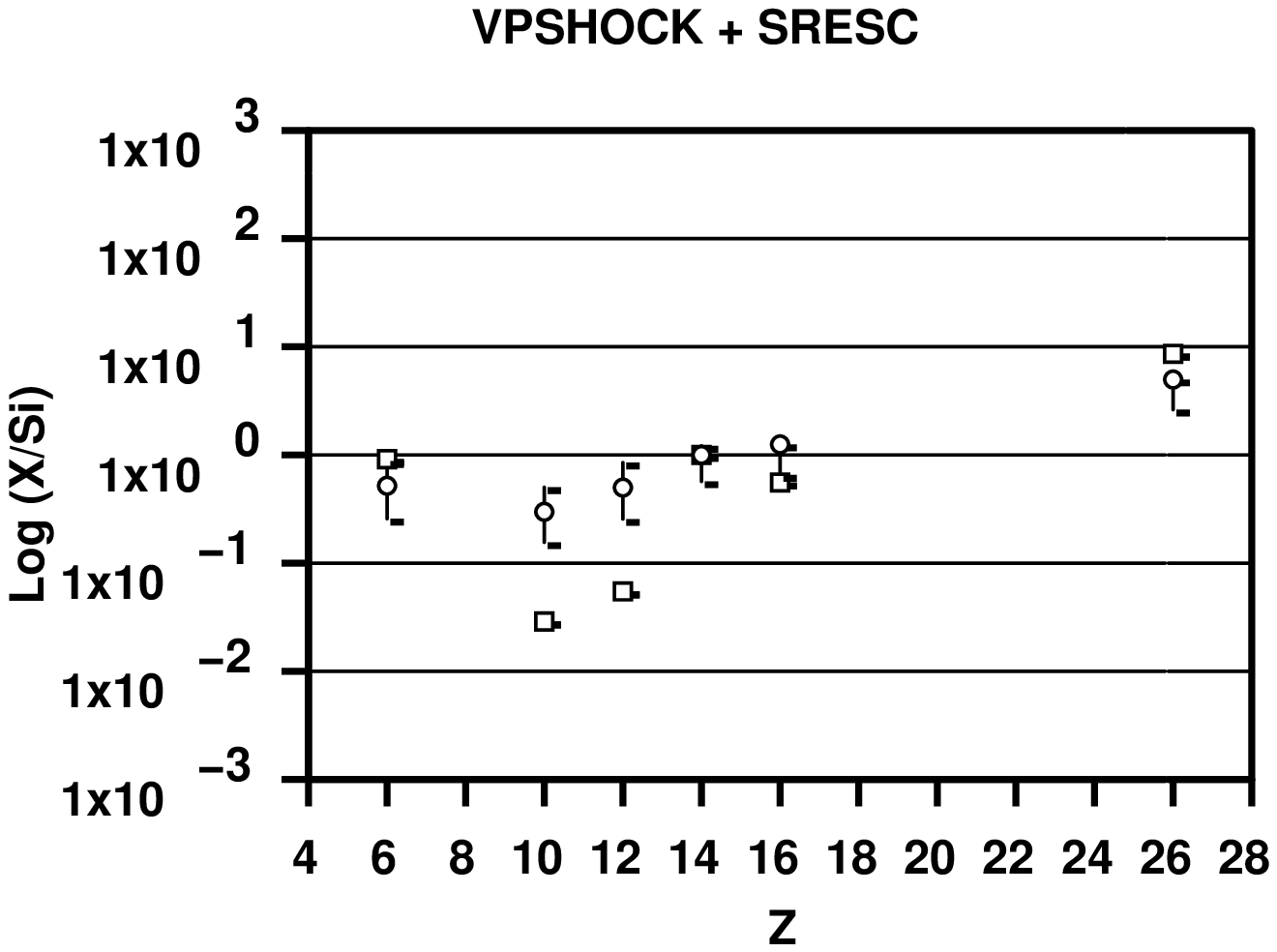} 
\caption{Abundances
measured by a) \vshock model (circles) and b) \vshock + \sresc (circles)
compared to the W7 supernova model from \citet{Iwamoto99} (squares),
normalized to silicon (by mass). \label{abund}} \end{figure}

\begin{figure} 
\plottwo{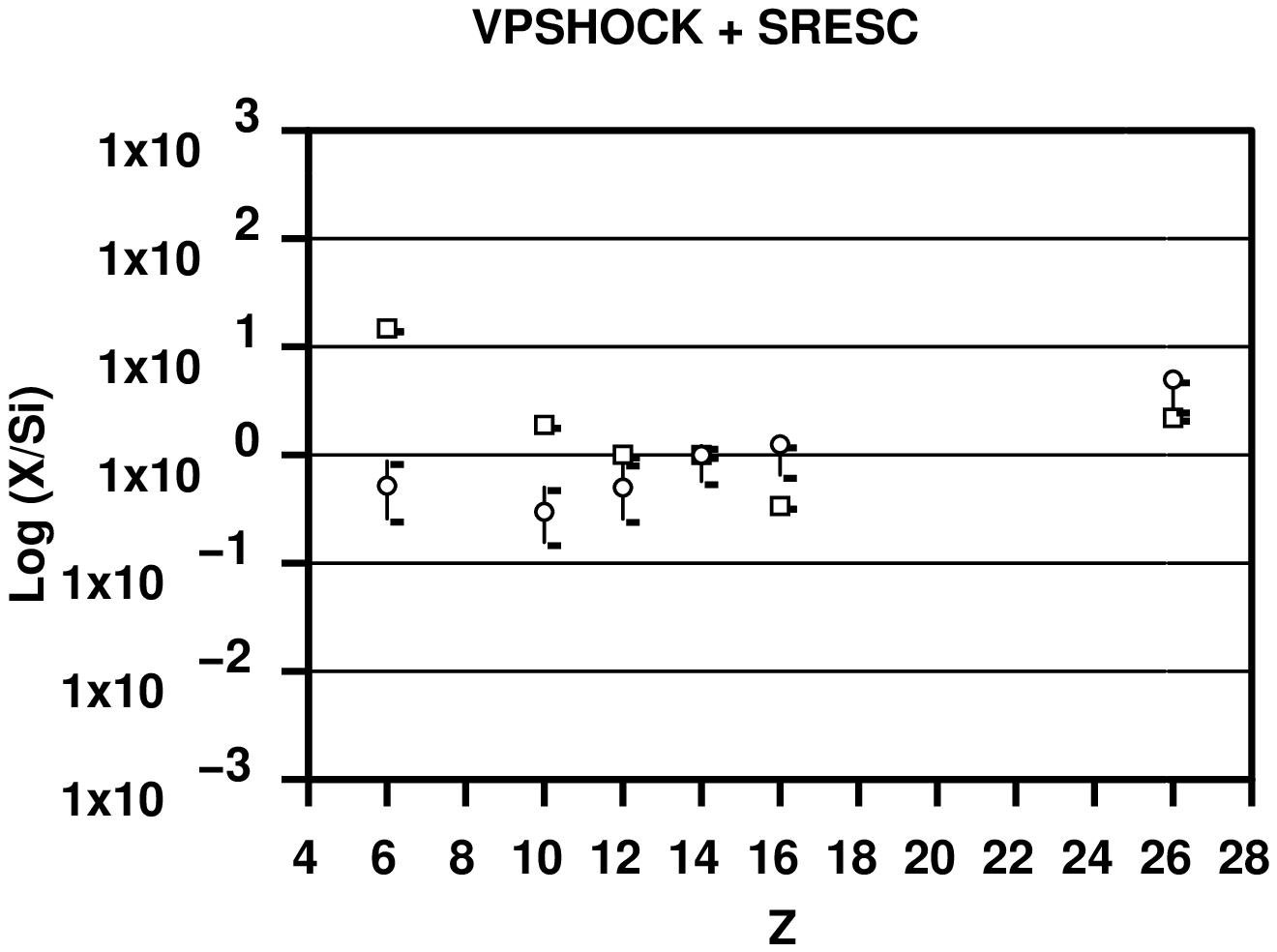}{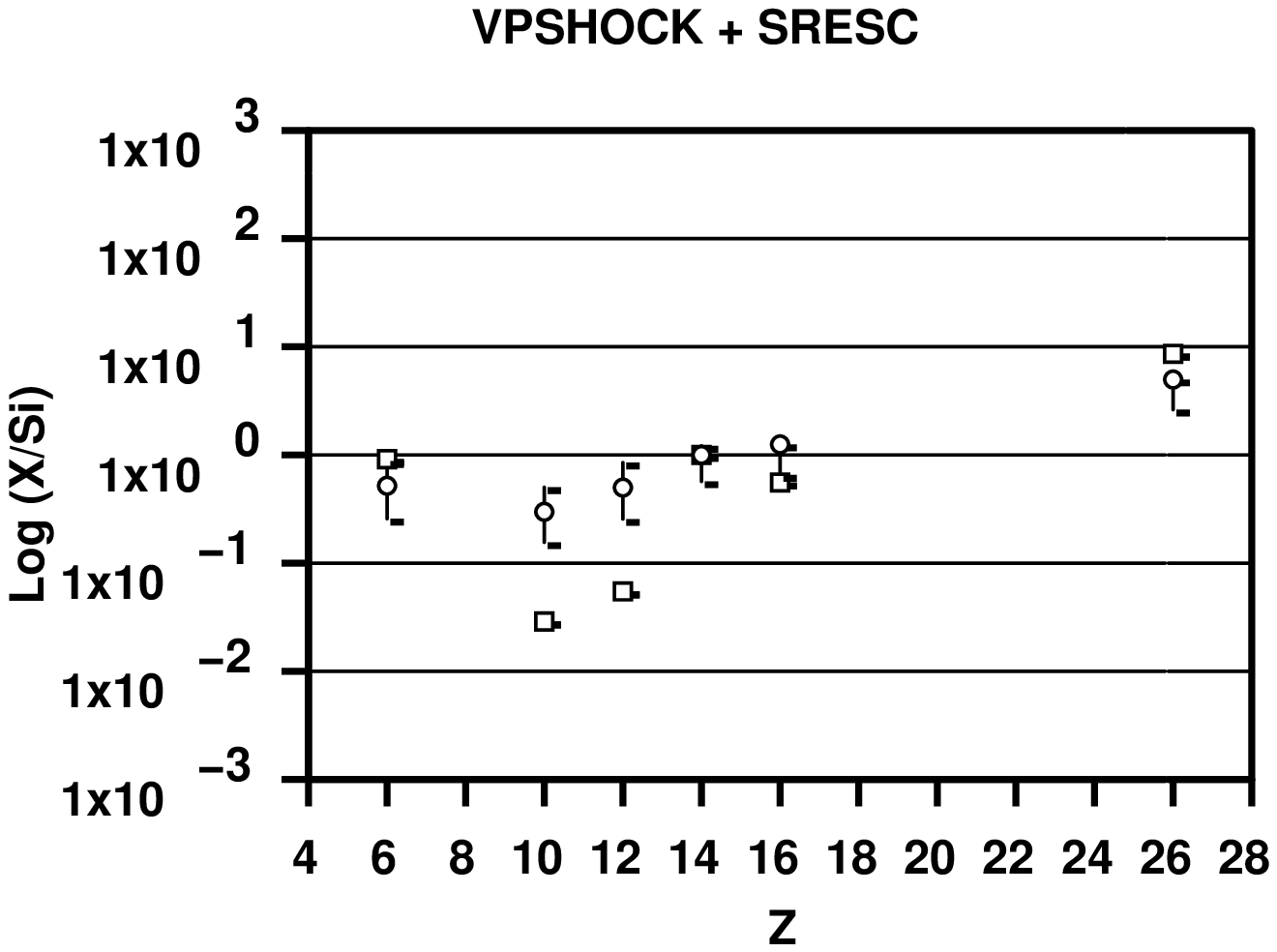} 
\caption{Abundances measured by
\vshock + \sresc (circles) compared to a) Type II SNR b) Type Ia (W7) SNR
predictions from \citet{Iwamoto99} (squares), normalized to silicon (by
mass). \label{abund2}} \end{figure}

In Figure \ref{abund} we compare the measured abundances \citep{Iwamoto99} with a thermal
model alone ({\it vpshock}), to abundances measured with a thermal + synchrotron model (\vshock+ {\it sresc}). The abundances are given relative to Si. The predicted abundance of Si for the W7 model is 0.16 M$_\sun$ and 0.12 M$_\sun$ for the Type II model \citep{Iwamoto99}.
The abundances in our fits change significantly with the addition of the nonthermal
model, bringing the measured abundances closer to models predicted by
\citet{Nomoto84}. We can take this farther and compare abundances
to those predicted for 
Type Ia and core-collapse SNRs. As shown in
Figure \ref{abund2}, the \vshock + \sresc model fit to the observations
seems to agree better with Type Ia models, as expected for SN~1006; however, the uncertainties are large and the Type II model cannot be ruled out.

By fixing silicon and fitting all other parameters, we are also able to calculate the total mass in each element, which we give in Table \ref{mass}. Most notably, with the \sresc model we have found half a solar mass of iron, closer to theoretical predictions, possibly a first for Type Ia SNR. As demonstrated in the table, while our calculated masses are not in perfect agreement with the Type Ia models in \citet{Iwamoto99}, they are closer to the Type Ia models than the Type II models.



We believe these results represent an improvement in technique over past work, while at the same time considering the values obtained preliminary. Improved results can be obtained by analyzing spatially resolved regions where the thermal emission dominates (such as will be done in Paper II). In addition we have used only the simplest possible thermal model (an single non-equilibrium ionization model not being physically plausible). 
However from these preliminary results we can state confidently that accurate accounting for a possible nonthermal component will be a necessary precondition to using nucleosynthesis model predictions to interpret abundances in SNRs.

\onecolumn

\subsection{\sresc inferences and cosmic-ray acceleration}

\subsubsection{Simple estimates}

Even before the reported detection of SN~1006 in TeV gamma-rays by the
\cangaroo\ collaboration \citep{Tanimori98}, Pohl (1996) calculated
expected TeV fluxes due to upscattering of cosmic microwave background
photons by electrons of 10-100 TeV, and the subsequent detection has
been widely accepted as direct evidence for the presence in SN~1006 of
such electrons.  Pohl calculated a range of values of predicted flux
based on a homogeneous emitting volume; the results depended on the
then poorly known electron spectrum.  

Since the maximum photon energy $\varepsilon_{\gamma}$ that can be
produced by an electron of Lorentz factor $\gamma_e$ upscattering photons
of energy $\varepsilon_s$ is given by $\varepsilon_{\gamma_e} = 4
\gamma_e^2 \varepsilon_s$, turning cosmic microwave background photons ($T
= 2.73$ K $\Rightarrow \epsilon_0 = 2.4 \times 10^{-4}$ eV) into 4 TeV
gamma-rays requires $\gamma \ge 6.5 \times 10^7$. Then the Klein-Nishina
parameter $\Gamma \equiv 4\gamma_e \varepsilon_s = 0.1$ and the Thomson
limit ($\Gamma < 1$) is marginally acceptable, at least for energies not
too far above the \cangaroo\ thresholds.  In that case, the inverse-Compton
volume emissivity for a thermal distribution of seed photons scattered by
an electron distribution $N_e(\gamma_e) = C \gamma_e^{-p}$ can be written
(Rybicki \& Lightman 1979, switching to photon frequencies for comparison
with synchrotron)
 \begin{equation} j_\nu = {h \over 4 \pi} {dn_\gamma
\over dt} =
        { {C (2 \pi r_0^2)} \over h^2 c^2} (kT)^{(p+5)/2} F(p)
        (h\nu)^{-(p-1)/2} \ {\rm erg \ cm}^{-3} \ {\rm s}^{-1}
	\ {\rm Hz}^{-1} \ {\rm sr}^{-1} \end{equation} where $F(p) = 5.67
(6.11) $ for $p=2.1 \ (2.2)$ (values appropriate for SN~1006).  Rewriting
the electron distribution as $N(E) = K E^{-p}$ with $K = C(m_e
c^2)^{p-1}$, and taking $p = 2.2$, we can express the ratio of the
synchrotron to inverse-Compton volume emissivities as \begin{equation}
{j_\nu ({\rm SR}) \over j_{\nu} ({\rm IC})} = {{3.48 \times 10^{-12} K
B^{1.6}_G \nu^{-0.6}} \over {2.35 \times 10^{-25} K \nu^{-0.6}}} = 3.72
\times 10^3 B_{\mu G}^{1.6} \end{equation} where $B_{\mu G}$ is the
magnetic field in the radiating region in units of $\mu G$.

We can use this result to estimate the mean magnetic field in SN~1006,
making the simplest assumptions of a homogeneous source coextensive in
synchrotron and inverse-Compton emission.  (The inverse-Compton emitting volume will
always be at least as large as the synchrotron volume, larger if electrons
diffuse to where the magnetic field is weak or negligible.) If no turnover
has begun by the lower of the two \cangaroo\ energies, the ratio of the
extrapolation of the synchrotron flux to the TeV flux at 1.7 TeV is about
$1.4 \times 10^5$, using a radio spectral index $\alpha = (p-1)/2 = 0.6$.  
This gives a mean magnetic field strength of 9.6 $\mu G$.

The observed radio flux density at 1 GHz of 19 Jy \citep{Green98} then
requires $K = 2.8 \times 10^{-12}$ erg$^{p-1}$ cm$^{-3}$.  Now the
immediate post-shock relativistic electron density is given by \begin{equation} u_e
\equiv \int_{E_l}^{E_h} E N(E) dE = {K \over {p - 2}}
    E_l^{2 - p} \left( 1 - \left( {E_h \over E_l}\right)^{2-p}\right)
    \cong 5 K E_l^{-0.2} \end{equation} where $E_h$ and $E_l$ are the ends
of the electron distribution, and we have assumed $E_h \gg E_l$.  If we
let $E_l = 0.5$ MeV, appropriate since a constant power-law in momentum
will flatten in energy below $mc^2$, we find a post-shock energy density
\begin{equation} u_e = 2.3 \times 10^{-10} \ {\rm erg \ cm}^{-3}
\end{equation} which can be compared with the post-shock pressure
\begin{equation} P_2 = {3 \over 4} \rho_1 u_{\rm sh}^2 = 9 \times 10^{-9}
    \left( {u_{\rm sh} \over {3200 \ {\rm km \ s}^{-1}}}\right)^2
\end{equation} for a strong shock in a gas with adiabatic index $5/3$.  
So this simple estimate implies a relativistic-electron acceleration
efficiency in the shock of about 2.6\%.  Extension of the same momentum
power-law to nonrelativistic energies would give a comparable amount
of energy in nonrelativistic suprathermal electrons. If the shock
is dominated by relativistic particles, the compression ratio could
be as high as 7, dropping the post-shock pressure by about 50\% and
increasing the inferred efficiency by that factor.

If supernova remnants are to produce the Galactic cosmic rays,
efficiencies of ion acceleration must be of order several tens of percent,
so given that we do not know the ratio of electron to ion acceleration
efficiency, these estimates seem reasonable.  They also imply a total
magnetic energy content in SN~1006 of \begin{equation} U_B \equiv {{4 \pi
R_{\rm sh}^3 \over 3} \phi {B^2 \over {8 \pi}}} = 5.3 \times 10^{46} \
{\rm erg} \end{equation} where $\phi$ is a volume filling factor of the
emission, taken to be $1/4$.  This is far below equipartition, since the
total relativistic-electron energy is \begin{equation} U_e = 3.3 \times
10^{48} \ {\rm erg} \end{equation} assuming the same filling factor, and
we have not taken account of ions.  However, if the pre-shock magnetic
field is of order 3 $\mu G$, the post-shock field due to shock compression
only will be about 10 $\mu G$, so reaching equipartition would require
major turbulent magnetic field amplification.

\subsubsection{Model calculations}

\begin{figure} 
\plotone{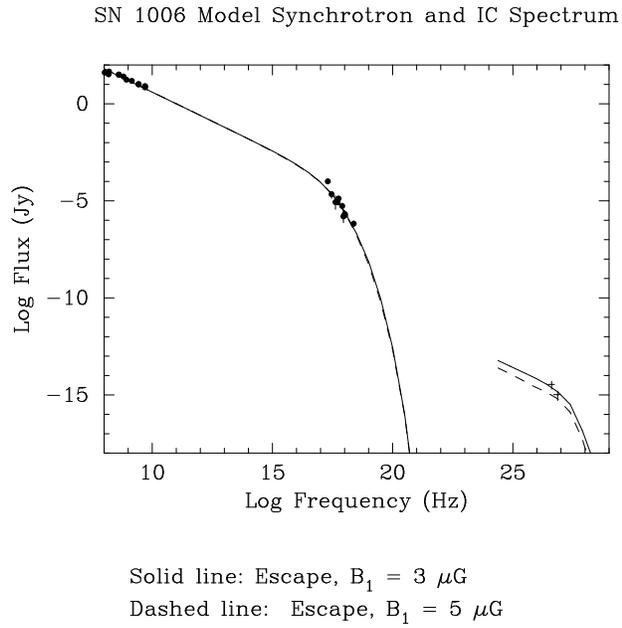}
\caption{Two gamma-ray spectra, for $B_1 = 3$ and $5 \
\mu$gauss ($r = 4$), constrained to have the same value of \norm, i.e., varying in
$K$. \label{gammaspectra}} \end{figure}

Using the full spectral code, we have calculated the spectrum shown in
Figure \ref{gammaspectra}, using the rolloff frequency value obtained from
our best fit result above.  Now fitting \sresc produces only the value of
\begin{equation} \nu_{\rm rolloff} \equiv 5.3 \ \nu_{max} = 5.3 \left(1.82
\times 10^{18} E_{max}^2 B_2 \right), \end{equation} where $B_2$ is the
post-shock magnetic-field strength and $E_{\rm max}$ is given by (1). (The
factor 5.3 was introduced in converting the original models into
XSPEC-readable format.) So 
\begin{equation} \nu_{\rm rolloff} = 5.57 \times 10^{15}
  \left({ \lambda_{\rm max} \over {10^{17} \ {\rm cm}}}\right)^2 B_{\mu G}^3
\left( r \over 4 \right).\end{equation} 
There is no theoretical expectation for $\lambda_{\rm max}$, so any
value of $B_1$ can produce a particular value of $\nu_{\rm rolloff}$. This
degeneracy is then broken by the inverse-Compton predictions which do
depend on $B_1.$

In Figure \ref{gammaspectra}, we show the model with our fitted value
of $\nu_{\rm rolloff} = 3.1 \times 10^{17}$ Hz, and $B_1 = 3 \ \mu G
\Rightarrow \lambda = 1.4 \times 10^{17} (4/r)^{-1/2}$ cm.  This model 
is able to
reproduce the \cangaroo\ fluxes reasonably well.  In the \sresc model,
the magnetic field behind the shock is compressed only, by a factor of
between 1 and $r$ depending on the shock obliquity, and drops with
distance behind the shock.  The mean value of $B_2$ is approximately
$9 \ \mu G$, consistent with the simple estimate above. The required
value of $K$ is $5.9 \times 10^{-12}$ erg$^{p - 1}$ cm$^{-3}$, about
twice the simple estimate, implying
\begin{equation} u_e = 4.8 \times 10^{-10} \ {\rm erg \ cm}^{-3}
\Rightarrow U_e = 7.0 \times 10^{48} \ {\rm erg} \end{equation} and a
relativistic-electron acceleration efficiency $u_e / (3/4) \rho_1 u_{\rm sh}^2 =
5.3$\% -- somewhat higher than the simple estimate. This value was not
obtained, as was the simple estimate, by assuming the Thomson limit.
Again, a somewhat higher compression ratio does not alter the electron
energy density, but by dropping the post-shock pressure slightly,
can increase this value by of order 50\%.

The values of $\lambda_{\rm max}$ and $B_1$ give $E_{\rm max} = 50$
ergs or 32 TeV from Equation (1); this energy is in the range of the upper
limits to maximum energies of synchrotron X-rays allowable in twelve
other Galactic remnants whose X-rays are dominantly thermal (and 
far below the ``knee'' at around 1000 TeV).  The
significance of this particular value of $\lambda_{\rm max}$, above
which the level of magnetohydrodynamic turbulence is presumed to be much less, is not
obvious; this is the first example of this form of determination of
structure in the magnetic-fluctuation power spectrum.

Figure \ref{gammaimage}b shows the predicted image in gamma rays at 1
TeV.  The modeling of the propagation of electrons escaping upstream
is not constrained at all by X-ray fitting, since those electrons
radiate in the weaker upstream magnetic field and hence contribute
only a small amount of synchrotron flux.  However, they contribute
more significantly to the inverse-Compton flux.  Imaging observations may allow
improvements in the description of these escaping electrons.  The
width of the emission behind the shock is greater than for synchrotron
X-rays, since the magnetic field also drops behind the shock. However,
in our nonthermal models for SN~1006, a contact discontinuity
separating shocked ISM from SN ejecta is still present (modeled by the
cutoff radius at which material was shocked at $10^4$ km s$^{-1}$),
and the shock-accelerated electrons are not assumed to diffuse into
that region.

There is no obvious reason that the NE limb, but not the SW, should be
detected in TeV gamma rays, especially since both limbs have very
similar X-ray spectra \citep{Allen97b}. \citet{Tanimori98}
report that the upper limit on emission from the SW limb is about
one-quarter of the emission detected from the NE.  This result demands
a lower relativistic-electron density by that factor. An unfortunate
coincidence could allow the magnetic-field strength to be larger in
the SW in such a way as to produce an identical synchrotron spectrum
with a lower density of relativistic electrons.  The magnetic field
would need to be larger by at least $4^{2/(p + 1)} = 2.4$.  Other
possible explanations are at least as contrived, such as greater
line-of-sight component of the magnetic field in the SW compensating
for a lower electron density. We are unable to offer any convincing physical 
explanation at this time for the detection of only the northeast limb.
We plan to search for spatial variations,
such as variations in the value of $\nu_{\rolloff}$, between the limbs,
which might help account for the lack of TeV emission from the SW, in
our spatially resolved study.


\section{Conclusions} 
\label{sec:conclusions} 

\begin{enumerate} 
\item The escape-limited synchrotron model, {\it sresc}, provides a good fit to
integrated spectral observations of SN~1006 by \rxte~ and \asca~ -- the highest quality data to date. In addition, the spatial prediction closely matches the X-ray image in the energy range where the model applies. 

\item The \sresc model provides a
significant improvement over the power-law models:

\begin{enumerate}
	
\item \sresc provides a more accurate description of the
	emission based on physical principles.
\item \sresc allows a
	clearer separation of thermal and nonthermal emission,
	constrained by radio observations 
\end{enumerate} 

\item We believe
an adequate description of the synchrotron emission leads to more accurate
temperature and abundance measurements in the remaining thermal model,
though better thermal inferences require both spatially resolved spectra
(and models) and codes appropriate for ejecta dominated by heavy elements.
Evidence for enhanced abundances suggests that at least part of the
thermal X-ray emission comes from ejecta.

\item The TeV gamma-ray observations allow us to determine the
energy in relativistic electrons to be about $7 \times 10^{48}$ erg; the
current shock efficiency at accelerating electrons is about 5\%, and
the energy in relativistic particles is much greater than that in magnetic
field.

\item We estimate that the ambient magnetic field is about 3 $\mu$Gauss, and
that the MHD wave spectrum near SN 1006 drops substantially in amplitude
above a wavelength of about $10^{17}$ cm.

\end{enumerate}

We have demonstrated that {\it sresc}, in conjunction with inhomogeneous
thermal models, can describe the full remnant emission.  These models 
(\sresc and \srcut in more general situations, as discussed in Section \ref{sec:model}) 
represent a significant 
improvement over power-law models in describing the physics of synchrotron X-ray emission. We can now
proceed to spatially-resolved spectral modeling of SN~1006. The true test
of the escape limited synchrotron model will be its application to the
spatially resolved data sets in our forthcoming paper. We will develop
specialized versions of the \sresc model to describe spatially distinct
areas of SN~1006.

\acknowledgments

The National Radio Astronomy Observatory is a facility of the National
Science Foundation operated under a cooperative agreement by Associated
Universities, Inc.  Our research made use of the following online
services: NASA's Astrophysics Data System Abstract Service, NASA's SkyView
facility (http://SkyView.GFSC.NASA.gov) located at NASA Goddard Space
Flight Center and SIMBAD at Centre de Donn\'ees astronomiques de
Strasbourg (US mirror http://simbad.harvard.edu/Simbad).

Thanks to J. Keohane for research notes and advice.  This research is
supported by NASA grant NAG5-7153 and NGT5-65 through the Graduate Student
Researchers Program.

\twocolumn

\clearpage
\onecolumn

\begin{deluxetable}{rcc} 
\footnotesize 
\tablecaption{\rxte \ Observations of SN~1006 AD \label{rxte}} 
\tablewidth{0pt} 
\tablehead{ 
\colhead{Parameter}&\colhead{\rxte \pca}} 
\startdata 
Cycle & A01 Epoch 1\\
Date & 1996 Feb 18-19 \\
Pointing center \al 	&15$^h$  04$^m$ 00\fs00\\
$\beta$ (J2000) & -41\arcdeg~48\arcmin~00\farcs0 \\
Exposure (ks) & 6.6 \\ 
Counts (k)  & 477 \\ 
Effective Resolution   & 1\arcdeg \ FWHM\\ 
Field of view  & 1\arcdeg\\ 
Energy range  &2-60 keV\\ 
\enddata 
\end{deluxetable}

\begin{deluxetable}{rccc} 
\footnotesize 
\tablecaption{Select \asca \ observations of SN~1006 AD \label{asca}} 
\tablewidth{0pt} 
\tablehead{ 
\colhead{Parameter}&\colhead{Center}&\colhead{Northeast}&\colhead{Southwest}}
\startdata 
Cycle 			&PV	&PV	&AO4\\
Date & 1993 Aug 19& 1993 September 13 & 1996 February 20 \\
Pointing center \al 	&15$^h$ 02$^m$ 48\fs24 & 15$^h$ 03$^m$ 31\fs92 &15$^h$ 02$^m$ 34\fs32\\
$\beta$ (J2000) &-41\arcdeg~55\arcmin~45\farcs8 & -41\arcdeg 
~46\arcmin~25\farcs0& -42\arcdeg~02\arcmin~57\farcs8\\
Exposure (ks)& 23 &23&22\\
\enddata 
\end{deluxetable}         

\begin{deluxetable}{rcccc} 
\footnotesize 
\tablecaption{Datasets Fit For This Paper \label{data}} 
\tablewidth{0pt} 
\tablehead{ 
&\multicolumn{2}{c}{{\bf North Limb}}&\multicolumn{2}{c}{{\bf Full Remnant}}\\
\colhead{Parameter}&\colhead{\asca \sis}&\colhead{\asca \gis}&\colhead{\asca \gis}&\colhead{\rxte \pca}} 

\startdata

Date & 1993 September 13&	1993 September 13 & 1993-1996& 1996 Feb 18-19 \\
Average Pointing center \al 	&15$^h$ 03$^m$ 55$^s$ &15$^h$ 03$^m$ 55$^s$&15$^h$ 02$^m$ 58$^s$&15$^h$  04$^m$ 00$^s$\\
$\beta$ (J2000) &-41\arcdeg~46\arcmin~00\arcsec &-41\arcdeg~46\arcmin~00\arcsec &-41\arcdeg~33\arcmin~02\arcsec&-41\arcdeg48\arcmin00\arcsec\\
Total Exposure (ks)	&104  	&104	&136	&6.6\\ 
Total Counts (k) 	&33 	&14	&258	&477\\
Background from 	& chip 	&night sky 	& night sky	&model\\
Background Counts (k) 	&4	&144	&9400	& 5.46 (from model)\\  
Spatial area used &$\sim$10\arcmin&$\sim$10\arcmin &$\sim$78\arcmin$\times$103\arcmin & $\sim$2\arcdeg \ \\ 
Energy range &0.6-10.0 keV &0.6-10.0 keV &0.6-10.0 keV & 3.0-10.0 keV\\ 
\enddata 
\end{deluxetable}       


\begin{deluxetable}{rcccc}
\footnotesize
\tablecaption{Model Fits to \asca \gis \& \rxte \pca observations of SN~1006 AD\label{fits}}
\tablewidth{0pt}
\tablehead{
\colhead{Parameters}&\colhead{\snei}&\colhead{\vshock}&\colhead{\vshock+\power}&\colhead{\vshock+\sresc}}
\startdata
$\chi^{2}$/DOF	& 7463/2285	& 3237/2279	&2721/2279	&2709/2280\\
$\chi_\nu^2$	& 3.27		& 1.42		&1.19		&1.19\\
\pca/\gis	&0.70		& 0.63		&0.64		&0.64\\
kT [keV]		& 1.63$_{1.64}^{1.62}$		&2.18$_{2.22}^{2.13}$		&1.79$_{1.74}^{1.84}$		&0.60$_{0.52}^{0.70}$\\
$\tau$ [s cm$^{-3}$]	& 7.3E8$_{7.18}^{7.42}$& 3.3E9$_{3.35}^{3.07}$	&2.9E9$_{2.72}^{3.14}$&	3.8E9$_{3.6}^{***}$\\
Abundances:\\
O		&1		&8.2E-2$_{7.0E-2}^{9.3E-2}$	&8.6E-2$_{7.4E-2}^{9.9E-1}$	&3.2E-2$_{2.0E-2}^{3.7E-2}$\\
Ne		&1		&2.8E-1$_{2.5E-1}^{3.1E-1}$	&2.5E-1$_{2.2E-1}^{2.9E-1}$	&1.0E-1$_{6.8E-2}^{1.2E-1}$\\
Mg		&1		&1.5E0$_{1.4E0}^{1.6E0}$	&1.2E0$_{1.1E0}^{1.4E0}$	&4.4E-1$_{3.0E-1}^{5.2E-1}$\\
Si		&1		&1.0E0$_{9.3E-1}^{1.1E0}$	&1.0E0$_{9.2E-1}^{1.1E0}$	&1.0E0$_{6.9E-1}^{***}$\\
S		&1		&5.3E-1$_{4.2E-1}^{6.6E-1}$	&7.4E-1$_{5.8E-1}^{9.3E-1}$	&2.2E0$_{1.5E0}^{2.5E0}$\\
Fe		&1		&1.9E0$_{1.6E0}^{2.4E0}$	&2.5E0$_{1.8E0}^{3.3E0}$	&2.1E0$_{1.4E0}^{2.8E0}$ \\
Normalization\tablenotemark{1}&	0.24$_{0.23}^{0.25}$ & 0.15$_{0.14}^{0.16}$&0.11$_{0.105}^{0.111}$/2E-2	&0.61$_{0.58}^{0.63}$/19.0\\
Nonthermal&\dots&\dots&$\Gamma$=2.50$_{2.34}^{2.63}$ keV$_{-1}$ cm$_{-2}$ s$_{-1}$	&\rolloff=3.0E17$_{2.8}^{3.1}$ Hz\\
Flux [ergs cm$^{-2}$ s$^{-1}$]	&2.0E-10		&1.9E-10		&1.3E-10/5.6E-11	&5.0E-11/1.3E-10\\
\tablenotetext{***}{In some cases it was not possible to establish errors on certain values. See Section \ref{sec:results} for discussion.}
\tablenotetext{1}{Normalization represents different physical quantities: For thermal models \norm~is the emission measure. For a powerlaw \norm~ is flux at 1 keV in photons keV$^{-1}$ cm$^{-2}$ s$^{-1}$. For \sresc~\norm~is the 1 GHz radio flux.}
\enddata
\end{deluxetable}

\begin{deluxetable}{rccc} 
\footnotesize 
\tablecaption{Mass of Elements in SN~1006 \label{mass}} 
\tablewidth{0pt} 
\tablehead{ 
\colhead{Element, [M$_\sun$]  }&\colhead{{\it vpshock}}&\colhead{W7}&\colhead{Core Collapse}} 
\startdata 
O 	&3.5E-2$_{2.7E-2}^{3.9E-2}$	&1.4E-1 &1.8E0\\
Ne 	&2.0E-2$_{1.8E-2}^{2.3E-2}$  	&4.5E-3 &2.3E-1\\
Mg 	&3.3E-2$_{3.0E-2}^{3.7E-2}$  	&8.6E-3 &1.2E-1\\
Si 	&8.1E-2$_{7.3E-2}^{1.6E-1}$ 	&1.6E-1 &1.2E-1\\
S 	&9.5E-2$_{7.7E-2}^{1.2E-1}$  	&8.7E-2 &4.1E-2\\
Fe 	&4.5E-1$_{3.9E-1}^{5.0E-1}$  	&7.5E-1 &9.1E-2\\
\enddata 
\end{deluxetable} 
\clearpage

\onecolumn

\clearpage

\begin{figure} 
\plottwo{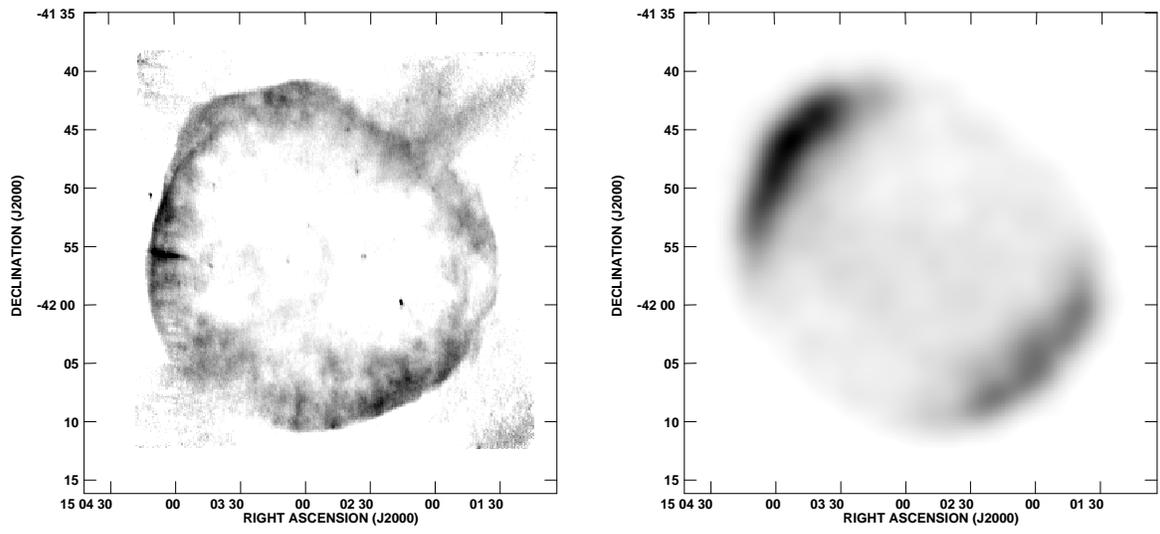}{f2b.ps} 
\caption{a) VLA image of
SN~1006 AD resolution 12\arcsec \ $\times$ 7\arcsec, corrected for the
primary beam. b) \asca \gis \ 3 convolved to 60\farcs \label{radioxray}}
\end{figure} 
\clearpage

\clearpage

\twocolumn

\begin{figure} 
\plotone{f4.ps} 
\caption{The \snei model fit to the
full remnant \asca \gis \ (diamonds) and \rxte \pca (crosses). Parameters listed in Table
\ref{data}, column 2.\label{snei}} \end{figure} 

 \clearpage

\onecolumn

\clearpage

\clearpage

\clearpage

\twocolumn

\begin{thebibliography}{DUM}

\bibitem[Aharonian \& Atoyan(1999)]{Aharonian99} Aharonian, F. A. \&
Atoyan, A. M. 1999, \aap, 351, 330

\bibitem[Allen et al.(1997b)]{Allen97b} Allen, G. E., Keohane, J. W.,
Gotthelf, E. V., Petre, R., Jahoda, K., Rothschild, R. E.,
Lingenfelter, R. E., Heindl, W. A., Marsden, D., Gruber, D. E.,
Pelling, M. R., \& Blanco, P. R. 1997b, ApJ, 487L, 97

\bibitem[Allen et al.(1997a)]{Allen97a} Allen,
G. E., Petre, R., Gotthelf, E. V. \& Keohane, J. 1997a, American
Astronomical Society Meeting, 191, 9903


\bibitem[Arnaud (1996)]{Arnaud} Arnaud, K. A., 1996, Astronomical Data Analysis
 Software and Systems V, eds. Jacoby G. \& Barnes J., p17, ASP Conf. Series volume 101. 

\bibitem[Asvarov et al.(1990)]{Asvarov90} Asvarov, A. I., Guseinov,
O. H., Kasumov, F. K., \& Dogel', V. A. 1990, A\&A, 229, 196

\bibitem[Baring et al.(1999)]{Baring99} Baring, M. G., Ellison,
D. C., Reynolds, S. P., Grenier, I. A., \& Goret, P. 1999, ApJ, 
513, 311

\bibitem[Becker et al.(1980)]{Becker80} Becker, R. H., 
Szymkowiak, A. E., Boldt, E. A., Holt, S. S. \& Serlemitsos, P. J. 1980, 
\apjl, 240, L33 

\bibitem[Berezhko et al. (1999)Berezhko, Ksenofontov, \& Petukhov]{Berezhko99} Berezhko, E. G., 
Ksenofontov, L. T., \& Petukhov, S. I. 1999, Proc.
26th Int.Cosmic-Ray Conf. (Salt Lake City), 4, 431

\bibitem[Berezhko et al.(1996)Berezhko Yelshin \& Ksenofontov]{Berezhko96}
Berezhko, E. G., Yelshin, V. K., \& Ksenofontov, L. T. 1996, 
Soviet Phys.JETP Lett., 82(1), 1

\bibitem[Blair et al.(1996)Blair, Long \& Raymond]{Blair96} Blair, W. P., 
Long, K. S. \& Raymond, J. C. 1996, \apj, 468, 871

\bibitem[Blumenthal \& Gould (1970)]{Blumenthal70} Blumenthal, G.\ 
R.\ \& Gould, R.\ J.\ 1970, Reviews of Modern Physics, 42, 237 

\bibitem[Borkoswki et al.(2000a)]{Borkowski00a}Borkowski, K. J., 
Rho, J., Dyer, K. K. \& Reynolds, S. P. 2000a, \apj, in press

\bibitem[Borkowski et al.(2000b)Borkowski, Lyerly \& Reynolds]{Borkowski00b} Borkowski, K. J., Lyerly, W. J. \& Reynolds, S. P. 2000b, submitted to \apj 

\bibitem[Borkowski et al.(1999)]{Borkowski99}
Borkowski, K. J., Rho, J., Dyer, K. K. \& Reynolds, S. P. 1999,
American Astronomical Society Meeting, 195, 4313

\bibitem[Dwarkadas \& Chevalier(1998)]{Dwarkadas98} Dwarkadas, 
V.\ V.\ \& Chevalier, R.\ A.\ 1998, \apj, 497, 807 

\bibitem[Ellison et al.(2000)Ellison, Berezhko, \& Baring]{Ellison00} 
Ellison, D. C., Berezhko, E. G., \& Baring, M. G. 2000, ApJ, 540, 292

\bibitem[Gaisser et al.(1998)Gaisser, Protheroe, \& Stanev]{Gaisser98}
Gaisser, T. K., Protheroe, R. J., \& Stanev, T. 1998, ApJ, 492, 219

\bibitem[Gorenstein(1975)]{Gorenstein75} Gorenstein, P. 1975, \apj, 
198, 95 

\bibitem[Green(1998)]{Green98} Green D. A. 1998, A Catalogue of
Galactic Supernova Remnants (1998 September version), Mullard Radio
Astronomy Observatory, Cambridge, United Kingdom (available on the
World-Wide-Web at ``http://www.mrao.cam.ac.uk/surveys/snrs/'').

\bibitem[Grevesse \& Anders(1989)]{Grevesse89} Grevesse, N.  \&
Anders, E.  1989, American Institute of Physics Conference Series,
183, 1


\bibitem[Hamilton et al.(1986)Hamilton, Sarazin, \& Szymkowiak]{Hamilton86}
Hamilton, A. J. S., Sarazin, C. L., \& Szymkowiak, A. E. 1986, 
ApJ, 300, 698



\bibitem[Iwamoto et al.(1999)]{Iwamoto99} Iwamoto, K. , Brachwitz,
F. , Nomoto, K. 'I. , Kishimoto, N. , Umeda, H. , Hix, W. R.  \&
Thielemann, F. -K.  1999, \apjs, 125, 439

\bibitem[Jahoda et al.(1996)]{Jahoda96} Jahoda, K. , Swank, J.  H.,
Giles, A. B., Stark, M. J., Strohmayer, T. , Zhang, W.  \& Morgan, E.
H. 1996, \procspie, 2808, 59


\bibitem[Koyama et al.(1995)]{Koyama95} Koyama, K., Petre, R.,
Gotthelf, E. V., Hwang, U., Matsura, M., Ozaki, M., Holt, \&
S. S. 1995, Nature 378, 255

\bibitem[Koyama et al.(1997)]{Koyama97} Koyama, K. , Kinugasa, K. ,
Matsuzaki, K. , Nishiuchi, M. , Sugizaki, M. , Torii, K. 'i. ,
Yamauchi, S.  \& Aschenbach, B.  1997, \pasj, 49, L7

\bibitem[Lagage \& Cesarsky(1983)]{Lagage83} Lagage, P. O. Cesarsky,
C. J. 1983, \aap, 125, 249

\bibitem[Laming(1998)]{Laming98} Laming, J. M.  1998, \apj, 499, 309


\bibitem[Mastichiadis \& de Jager (1996)]{MastdeJ96} Mastichiadis, 
A., \& de Jager, O. C. 1996, \aap, 311, L5

\bibitem[Muraishi et al.(2000)]{Muraishi00} Muraishi, H., et al. 
2000, \aap, 354, L57 

\bibitem[Nomoto et al.(1984)Nomoto, Thielemann, \& Yokoi]{Nomoto84}
Nomoto, K., Thielemann, F. -K., Yokoi, K. 1984 \apj 286, 644


\bibitem[Pohl(1996)]{Pohl96} Pohl, M. 1996, A\&A, 307, 57

\bibitem[Reynolds \& Chevalier(1981)]{Reynolds81} Reynolds, S. P.  \&
Chevalier, R. A. 1981, \apj, 245, 912

\bibitem[Reynolds(1996)]{Reynolds96} Reynolds, S. P. 1996, \apjl, 459,
L13

\bibitem[Reynolds(1997)]{Reynolds97} Reynolds, S. P. 1997, BullAAS,
29, 1267


\bibitem[Reynolds(1998)]{Reynolds98} Reynolds, S. P. 1998, \apj, 493,
375

\bibitem[Reynolds(1999)]{Reynolds99b} Reynolds, S. P. 1999, 
Proc.~3rd INTEGRAL Workshop: Astr.Lett.\&Comm., 38, 425

\bibitem[Reynolds \& Keohane(1999)]{Reynolds99} Reynolds, S. P. \&
Keohane, J. W. 1999, \apj, 525, 368

\bibitem[Reynolds \& Gilmore(1986)]{Reynolds86} Reynolds, S.\ 
P.\ \& Gilmore, D.\ M.\ 1986, \aj, 92, 1138 


\bibitem[Rybicki \& Lightman (1979)]{Rybicki79} Rybicki, G.\ B.\, 
\& Lightman, A.\ P.\ 1979, New York, Wiley-Interscience, 1979.\ 393 p., 

\bibitem[Slane et al.(1999)]{Slane99} Slane, P. , Gaensler, B.  M.,
Dame, T. M., Hughes, J. P., Plucinsky, P. P., \& Green, A.  1999, \apj,
525, 357

\bibitem[Sturner et al.(1997)]{Sturner97}Sturner, S. J., Skibo, J. G.,
Dermer, C. D., \& Mattox, J. R. 1997, ApJ, 490, 619

\bibitem[Tanimori et al.(1998)]{Tanimori98}Tanimori, T., et al.~1998,
ApJ, 497, L25

\bibitem[The et al.(1996)]{The96} The, L. -S., Leising, M.  D.,
Kurfess, J. D., Johnson, W. N., Hartmann, D. H., Gehrels, N., Grove,
J.  E. \& Purcell, W. R. 1996, \aaps, 120, C357

\bibitem[Willingale et al.(1996)]{Willingale96}
Willingale, R., West, R. G., Pye, J. P. \& Stewart, G. C. 1996,
\mnras, 278, 749

\bibitem[Vink et al.(1997)Vink, Kaastra \& Bleeker]{Vink97} Vink, J. , Kaastra,
J. S. \& Bleeker, J. A. M. 1997, \aap, 328, 628

\bibitem[Vink et al.(2000)]{Vink00} Vink, J., Kaastra, J.\ S., Bleeker, J.\ A.\ M.\ \& Preite-Martinez, A.\ 2000, \aap, 354, 931 


\end{thebibliography}
\end{document}